\documentclass[sigconf, nonacm]{acmart}

\usepackage{caption}
\usepackage{scalerel}
\usepackage{graphicx}
\usepackage{xcolor}
\usepackage{tcolorbox}
\usepackage{amsmath,amsfonts}
\usepackage{ifsym}
\makeatletter
\let\c@lofdepth\relax
\let\c@lotdepth\relax
\makeatother
\usepackage{subfigure}
\tcbset{before={\par\pagebreak[0]\smallskip\parindent=0pt},after={\par\noindent}}
\usepackage{tree-dvips}
\usepackage{multirow}

\usepackage{qtree}
\newcommand\vldbdoi{XX.XX/XXX.XX}
\newcommand\vldbpages{XXX-XXX}
\usepackage[ruled,linesnumbered,vlined]{algorithm2e}
\usepackage{algorithmic}
\usepackage{textcomp}
\newcommand\vldbvolume{14}
\newcommand\vldbissue{1}
\newcommand\vldbyear{2022}
\newcommand\vldbauthors{\authors}
\newcommand\vldbtitle{\shorttitle} 
\usepackage{tikz}
\newcommand*\circled[1]{\tikz[baseline=(char.base)]{
    \node[shape=circle,black,draw,inner sep=0.5pt] (char) {#1};}}
\newcommand\vldbavailabilityurl{URL_TO_YOUR_ARTIFACTS}
\newcommand\vldbpagestyle{plain} 
\usepackage{listings} 
\newcommand{\yiming}[1]{\textcolor{blue}{#1}}

\newcommand{\squishlist}{
	\begin{list}{$\bullet$}
		{
			\setlength{\itemsep}{0pt}
			\setlength{\parsep}{1pt}
			\setlength{\topsep}{1pt}
			\setlength{\partopsep}{0pt}
			\setlength{\leftmargin}{1.5em}
			\setlength{\labelwidth}{1em}
			\setlength{\labelsep}{0.5em} } }
	
\newcommand{\squishend}{\end{list}}
\usepackage{multirow}

\begin{document}
\title{QUIP: Query-driven Missing Value Imputation}

\author{Yiming Lin}
\affiliation{%
  \institution{University of California, Irvine}
}
\email{yiminl18@uci.edu}

\author{Sharad Mehrotra}
\affiliation{%
  \institution{University of California, Irvine}
}
\email{sharad@ics.uci.edu}

\begin{abstract}
This paper develops a query-time missing
value imputation framework, entitled QUIP, 
that minimizes the joint costs of  imputation and query execution. 
QUIP achieves this by modifying how 
relational operators are processed. 
It adds a cost-based decision function in each operator
that checks whether the operator  
should invoke imputation prior to execution or 
to defer the imputations for downstream operators to resolve. 
QUIP implements a new approach to evaluating outer join that  preserve missing values during query processing,  and a bloom filter based index structure to optimize the space and running overhead.
We have implemented QUIP using ImputeDB - a specialized database
engine for data cleaning. 
Extensive experiments on both real and synthetic data sets demonstrates the effectiveness and efficiency of QUIP, which outperforms the state-of-the-art ImputeDB by 2 to 10 times on different query sets and data sets, and achieves the order-of-magnitudes improvement over offline approach.  

\end{abstract}

\maketitle

\pagestyle{\vldbpagestyle}
\begingroup\small\noindent\raggedright\textbf{PVLDB Reference Format:}\\
\vldbauthors. \vldbtitle. PVLDB, \vldbvolume(\vldbissue): \vldbpages, \vldbyear.\\
\href{https://doi.org/\vldbdoi}{doi:\vldbdoi}
\endgroup
\begingroup
\renewcommand\thefootnote{}\footnote{\noindent
This work is licensed under the Creative Commons BY-NC-ND 4.0 International License. Visit \url{https://creativecommons.org/licenses/by-nc-nd/4.0/} to view a copy of this license. For any use beyond those covered by this license, obtain permission by emailing \href{mailto:info@vldb.org}{info@vldb.org}. Copyright is held by the owner/author(s). Publication rights licensed to the VLDB Endowment. \\
\raggedright Proceedings of the VLDB Endowment, Vol. \vldbvolume, No. \vldbissue\ %
ISSN 2150-8097. \\
\href{https://doi.org/\vldbdoi}{doi:\vldbdoi} \\
}\addtocounter{footnote}{-1}\endgroup

\ifdefempty{\vldbavailabilityurl}{}{
\vspace{.3cm}
\begingroup\small\noindent\raggedright\textbf{PVLDB Artifact Availability:}\\
The source code, data, and/or other artifacts have been made available at \url{\vldbavailabilityurl}.
\endgroup
}

\section{Introduction}
\label{sec:introduction}

\subsection{Background}

A large number of real-world datasets contain missing values.
Reasons include human/machine errors in data entry, unmatched columns in data integration~\cite{kim2003taxonomy}, etc. Failure to clean the missing data may result in the poor  quality of answers to queries that may, in turn, negatively influence tasks such as machine learning~\cite{li2021cleanml}, data analytics, summarization~\cite{graham2009missing,herschel2009artemis}, etc. built on top of  data. 

Missing value imputation has been extensively studied in the literature, especially from the perspective of ensuring accuracy~\cite{lin2020missing,young2011survey,song2015enriching,burgette2010multiple}. 
Traditional Extract, Transform and Load (ETL) data processing pipeline treats missing value imputation as part of an  offline data preparation process that
cleans all the data prior to making it available for analysis. Such an data preparation step is often 
costly if data is large and if cost per imputation is high.

With the recent drive towards Hybrid transaction/analytical processing (HTAP) systems~\cite{athanassoulis2019optimal}
(that support real-time analysis on data as it arrives) and emergence of the extract-load-transform (ELT) process (wherein data is loaded into data lakes and transformed, e.g., cleaned,  at a later time),  we need to rethink how data cleaning 
should be performed in preparing data for analysis. 
In particular, database systems should support mechanisms to clean data  lazily, when the need to clean the data arises during query processing.  Compared to cleaning the entire dataset, cleaning only the portion needed for the query during query processing reduces the wasted effort and computational resources. Furthermore, cleaning the entire data set might not even be feasible in  domains where datasets can be very large. Often, predicting the right dataset to clean is not feasible, specially when analysts may perform ad-hoc queries on the data. 
In such situations, the only recourse  is to clean data needed to answer the query during query processing. 

Query-time data cleaning 
 opens the challenge of  co-optimizing data  cleaning with query processing. Since  
 data cleaning as part of a query increases query latency, 
 technique must be developed to  minimize data
 that is cleaned in order to answer
 the query so as to reduce  latency. 
Query-time data cleaning approaches  have been previously studied for a variety of data cleaning tasks. For instance, ~\cite{altwaijry2015query,altwaijry2013query} developed query-time cleaning approaches for entity resolution  while ~\cite{giannakopoulou2020cleaning} develops a query-time  cleaning approach for violation detection  and repair. 
ImputeDB~\cite{cambronero2017query} develops a method for query-time missing value imputation  (which is also the problem we study in this paper) and refer to it as  \textit{on-the-fly} imputation. We note that the term "On-the-fly" data cleaning has also been used in a different sense in some other works such as as ~\cite{ioannou2010fly,bilenko2005adaptive} where the goal is to clean the entire database but to make data  available iteratively while the cleaning is in progress \footnote{Such a problem is complementary to the lazy cleaning approach at query time and should not be confused with query-time approaches.}.

\begin{table*}
\small
                \begin{minipage}{0.33\textwidth}
                \centering
                \begin{tabular}{|c|c|c|}
                    \hline
                    Mac$\_$address & Time & Room$\_$location \\ \hline
                    4fep & 12pm & 2206  \\ \hline
                    3a4b & 2pm & NULL (\textcolor{blue}{$N_1=3001$}) \\ \hline
                    4fep & 1pm & NULL (\textcolor{blue}{$N_2=2082$}) \\ \hline 
                    25ya & 3pm & NULL (\textcolor{blue}{$N_3=2099$})\\ \hline   
                    fff1 & 1pm & 3119 \\
                    \hline
                    9aa4 & 2pm & 2214 \\ 
                    \hline
                \end{tabular}
                \caption{Trajectories (T)}
                \label{tab:T}
                \end{minipage}
                \begin{minipage}[c]{0.33\textwidth}
                \centering
                \begin{tabular}{|c|c|c|}
                    \hline
                    Room & Floor & Building\\ \hline
                    2214 & 2 & NULL (\textcolor{blue}{$N_4=$DBH}) \\ \hline
                    2206 & NULL (\textcolor{blue}{$N_5 = $2}) & DBH\\ \hline
                    2011 & 2 & DBH \\ \hline   
                    3119 & 3 & NULL (\textcolor{blue}{$N_6 = $ICS}) \\ \hline
                    2065 & 2 & NULL (\textcolor{blue}{$N_7 = $DBH}) \\ \hline   
                \end{tabular}
                \caption{Space (S)}
                \label{tab:S}
                \end{minipage}%
                \begin{minipage}[c]{0.33\textwidth}
                \centering
                \begin{tabular}{|c|c|c|}
                    \hline
                    Name & Email & Mac$\_$address\\ \hline
                    Mike & mike@uci.edu & NULL (\textcolor{blue}{$N_8 = $fff1}) \\ \hline
                    Robert & robert@uci.edu & 4fep\\ \hline
                    John & john7@uci.edu & NULL (\textcolor{blue}{$N_9=$9aa4}) \\ \hline
                \end{tabular}
                \caption{User (U)}
                \label{tab:U}
                \end{minipage}
                \vspace{-3em}
\end{table*}

\subsection{QUIP}
This paper  develops \textit{QUIP}, a \textit{QU}ery-time approach for m\textit{I}ssing value im\textit{P}utation that exploits query semantics to reduce the cleaning overhead.  Specifically, given as input an SQL query 
and  a corresponding query plan generated by any commercial relational query optimizer (e.g., such as PostgreSQL) or a specialized optimizer  designed to reduce imputation costs (e.g., such as ImputeDB~\cite{cambronero2017query}), QUIP develops an execution strategy that minimizes the overall (joint) cost of imputing missing data and executing the query based on the query plan.  

\begin{figure}
\raggedright
\texttt{
\hspace{-0.5em}\textcolor{blue}{SELECT} U.name, T.time, T. room$\_$location \\ \textcolor{blue}{FROM} Trajectories as T, Space as S, User as U \\
\textcolor{blue}{WHERE}
T.mac$\_$address = U.mac$\_$address
\textcolor{blue}{AND} \\
T.Room$\_$location = S.room
\textcolor{blue}{AND}
S.building = ‘DBH’ 
\textcolor{blue}{AND} \\
T.Room$\_$location in \{2065, 2011, 2082, 2035, 2206\}
}
\vspace{-1em}
    \caption{Query}
    \label{fig:query}
    \vspace{-2em}
\end{figure}

We illustrate the key insights behind QUIP by using a motivating example. 
Consider a real data set, WiFi location data, collected by sensor data management system  \textit{TIPPERS}~\cite{mehrotra2016tippers} in UCI. In Tables~\ref{tab:T}-\ref{tab:U}, we select a subset of WiFi location data,  \textit{Trajectories}, \textit{Space} and \textit{User}.  \textit{Trajectories} collects the room locations of devices with the corresponding time stamps, \textit{Space} stores the metadata for room, floor and building where this room is located, and \textit{User} records the user information that  contains the name, email of the user and mac address of the device owned by this user. 
There are totally 9 missing values in Table~\ref{tab:T},~\ref{tab:S} and ~\ref{tab:U} (represented as NULL values, from $N_1$ to $N_9$, and the blue colored values in brackets are real values that would result were we impute the missing values ), and we treat the imputation method to repair such missing values as black box. 

Now consider a commonly used SPJ query in our scenario, in Figure~\ref{fig:query}, that computes the trajectories (i.e., time and room  location) for all registered users (i.e., users in \textit{User} table) in the locations of interest (e.g., specific rooms in Donald Bren Hall (DBH) building). To reduce number of unnecessary imputations, 
query execution could delay imputing a missing value 
until the tuple (with the missing value) encounters
an  operator that
cannot be evaluated without imputation. 
Such an approach prevents cleaning data 
that is not necessary to answer the query, and has been used in systems such as ~\cite{altwaijry2013query}
and in ImputeDB~\cite{cambronero2017query}  in the context of missing value imputation. While delaying imputation (until an operators that require data to be cleaned is encountered)
reduces the overhead, nonetheless, as we will show,  much further improvement could be possible by modifying operator implementations to 
preserve tuples with missing values thereby enabling the possibility to further delay
 imputation to later in the query execution. 
 Such a delay could prevent imputation if the tuple
 containing the missing value is discarded
 by  downstream operators based on different predicates
 whose evaluation does not require missing value to be imputed. Let us see such a possibility of saving
 imputations using an example.

Consider first  the strategy that  cleans data as soon as it encounters an operator that requires it to be cleaned (as in ImputeDB). The total imputation cost will depend upon the query plan chosen to execute the query. 
For the query in Figure~\ref{fig:query}, several different query plans are possible.  Consider, for instance, a plan where selections (on both tables $T$ and $S$) are pushed
before the join.  Irrespective of the join order,  all  missing values for  attributes, i.e., \textsf{T.room\_location},  \textsf{S.Building}, and those under join attributes, i.e., \textsf{U.mac\_address}, will be imputed  resulting in a total of   \textit{8} imputations. An alternate plan might be to delay selections after the joins with the cost of potential high query processing overhead.  
If, for instance, \textsf{T.room\_location=S.Room}  executes first, 
3 imputations will be invoked  to impute the missing values for the  \textsf{T.room\_location} ($N_1,N_2,N_3$). It will result in 
3 matched tuples (with room location as 2206, 3119, 2214). Next $N_8,N_9$ will need to be imputed to execute $T\bowtie U$, which returns further 3 matches (with mac address as 4fep, fff1 and 9aa4). Now it requires 2 further imputations ($N_4, N_6$) to evaluate selection predicate \textsf{S.Building = 'DBH'}. 
Thus, such a plan would result in \textit{7} imputations. 
We note that  a plan where  \textsf{T.Room$\_$location=S.room} is executed first also needs \textit{5} imputations. Finally, if we only delay one of the two selections, and clean as soon as an operator encounters missing value, we need  5 (or 6) imputations depending on which selection predicate is pushed down/pulled up. 

In contrast, let us now consider a strategy that  
has the capability to delay imputations and preserve missing values in query processing. Such a strategy could  delay imputing prior to selection operations, i.e., \textsf{S.Building = 'DBH'} and \textsf{T.Room\_location = \{2065, 2011, 2082, 2035, 2206\}}, and join operation \textsf{T.Room\_location = S.Room} and impute  missing values \textsf{U.mac\_address} ($2$ missing values), when executing the join operation \textsf{T.mac\_address = U.mac\_address}. This will result in 2 matches (mac$\_$address = 4fep) and the other tuples will be eliminated. Then such a strategy will impute $N_2$ which is equal to 2082, leading to no  match in Space table and thus computing the query above with just 3 imputations. 

While this is a  toy example with modest savings of a few
 imputations,   in real data set, where cardinality of Trajectories and User table are large (e.g.,  over 40 million trajectories tuples and 60k+ devices (users) tuples in 10 months UCI-WiFi data set), such an approach can provide an order of magnitude reduction in the number of imputations. With 
the imputation being expensive 
such savings can be very significant. 
Such savings in imputations,  as will become
clear later in the paper, do come with the increased overhead of preserving  tuples with missing values in operators. The key is to design a solution that minimizes the overhead paid by operators to preserve such tuples, and to design a cost-based approach to judiciously decide whether to preserve tuples (and hence potentially reduce imputation costs) or to impute right away so as not to incur overhead of preserving tuples.

This paper develops QUIP,  a lazy approach to
imputing missing values during query processing,   that decouples the need to impute data from the logic of operator implementation.
QUIP makes two main modifications to existing query processing. First, it modifies current implementations of the relational operators to be aware that the data may  contain missing values and extends the operator implementations to preserve
such tuples.  For example, consider the selection condition  \textsf{S.building = 'DBH'} in the query shown in Figure~\ref{fig:query}. 
If the selection operator sees a missing value $v$, it decides on whether to impute $v$ and evaluate the predicate precisely, or to defer imputation (and correspondingly selection condition evaluation) to downstream operators to handle. 
The benefit of delaying arises if the downstream operator eliminates the object (due to its other associated predicate) thus preventing the need to impute
the missing values $v$. QUIP develops efficient ways to implement preserving tuples with missing values without
the overhead of quadratic increase in size as would be the case if join conditions were naively extended to preserve
missing values. Instead,  QUIP uses  a carefully designed outer join  for this purpose. 

With operators extended to preserve tuples, QUIP successfully decouples imputation from operator implementation allowing
for  missing value imputations to be incorporated anywhere in the query tree. Such a capability allows
QUIP to reduce the number of imputations in the example above. QUIP can delay imputing prior to selection operations, i.e., \textsf{S.Building = 'DBH'} and \textsf{T.Room\_location = \{2065, 2011, 2082, 2035, 2206\}}, and join operation \textsf{T.Room\_location = S.Room} and, instead, impute  missing values \textsf{U.mac\_address} ($2$ missing values), when executing the join operation \textsf{T.mac\_address = U.mac\_address}. This will result in 2 matches (mac$\_$address = 4fep) and the other tuples will be eliminated.  QUIP, then,  imputes $N_2$ which is equal to 2082, leading to no  match in Space table. Thus,  QUIP can process the query with the (minimal) number of imputations - viz., 3 in this case.

Note that, QUIP's goal is not just to minimize  imputations. Instead, it minimizes the  joint cost of imputing and query processing. If the imputation is cheap, QUIP will tend to first impute missing values instead of delaying them to reduce query execution overhead.  To achieve such a goal, the 
 second modification QUIP makes is to  use a cost-based approach to automatically guide each operator on whether to impute or delay missing values by considering both imputation and query processing overheads.

\noindent\textbf{Contribution:} 
The paper introduces QUIP framework to answer SQL queries over data that may contain missing values. QUIP judiciously delays imputations to minimize the combined cost of imputing and query processing simultaneously. The key to QUIP is a  time and space-efficient outer-join based mechanism to preserve missing values as well as several efficient data structures to wisely remove unnecessary missing values in query processing. QUIP  outperforms the state-of-the-art solutions by 2 to 10 times on different query sets and data sets, and achieves the order-of-magnitudes improvement over offline approach. 
In the rest of the paper,  
in Section~\ref{sec:relatedwork} and ~\ref{sec:problem}, we described the preliminaries and overview of the approach.  Section~\ref{sec:structure} introduces data structures used in QUIP. Section~\ref{sec:algorithm} to ~\ref{sec:decide}  
describe QUIP algorithm. Section~\ref{sec:evaluation} evaluates QUIP, and  Section~\ref{sec:conclusion} concludes the paper. 

\vspace{-0.5em}
\section{Preliminaries}
\label{sec:relatedwork}

\vspace{-0.25em}
\subsection{Imputation Operation}
 As in ~\cite{cambronero2017query},  we view   imputation approaches
as  \textit{blocking} or \textit{non-blocking}  in terms of query processing. A blocking strategy  reads the whole data to learn a model for imputation (before it imputes any missing value), while  
a non-blocking strategy  can impute
missing values independently reading only a (subset of related)  tuples.
Imputation approaches  can roughly be characterized as statistics based, rule based, master data based,  time-series based, or learning based approaches~\cite{lin2020missing}.
 Of these, other than
the learning based approaches, many techniques could be used in a non-blocking setting.
For instance, ImputeDB~\cite{cambronero2017query} used a non-blocking statistics-based
mean-value method that replaces a missing value with the mean  of the available values in the same column using histograms. Since  histograms are often maintained  for query optimization and approximate processing~\cite{thaper2002dynamic,jagadish1998optimal} such a technique is non-blocking. 
Strategies that use master data  ~\cite{chu2015katara,ye2016crowdsourcing,ye2020effective,qi2018frog} are also
non-blocking since they  look up a   knowledge base and crowd source the imputations one tuple (or a set of tuples) at a time. Imputation strategies in time series data ~\cite{khayati2020mind,bansal2021missing,lin2020locater}  are
often performed by learning patterns over \textit{historical} data to forecast \textit{current} missing values or using the correlation across the time series. 
An example  is LOCATER~\cite{lin2020locater}  that imputes each missing
location of a user at one time stamp by learning user's pattern from historical data. 
Such methods also clean one tuple at a time and are, hence, non-blocking.
Rule based imputation methods based on differential dependency   ~\cite{song2015enriching} or editing rules~\cite{fan2012towards}  often 
   impute missing values by replacing them with corresponding value of similar objects. 
 
 In non-blocking strategies the overall cost of imputation is proportional to the number of tuples imputed and hence, QUIP, 
 which is designed to exploit query semantics to reduce number of imputations
performed, can bring  significant improvements.

\begin{figure}[tb]
    \centering
    \includegraphics[width=6.5cm,height=4.3cm]{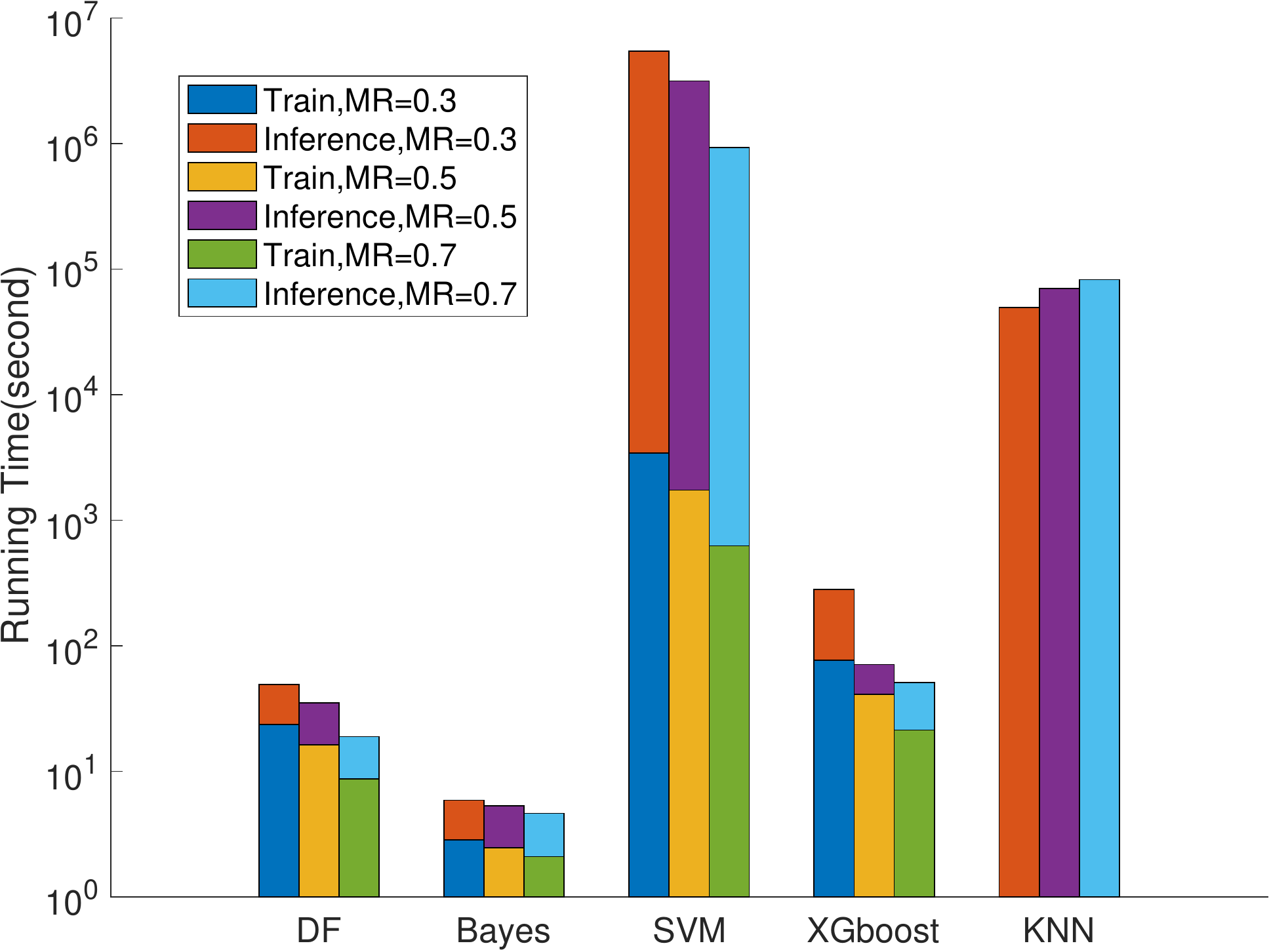}
    \vspace{-1.3em}
    \caption{Training and Inference time for different MLs.}
    \vspace{-2em}
    \label{fig:ML}
\end{figure}

In contrast to the above, learning-based approaches are often blocking. Such techniques, 
as in  \cite{miao2021efficient}, 
may use models such as generative adversarial network (GAN) ~\cite{arjovsky2017wasserstein} that can take over hours to train. Such techniques are, hence,  not suitable for being called from within online analysis queries due to long latency due to training prior to imputation. 
Given such a  limitation, several
prior works have explored reducing the training time of learning-based methods. 
Miao et al~\cite{miao2021efficient} select a small representative sample 
(about 5\%) to speed training by about 4x while maintaining imputation accuracy guarantees. Two widely used learning approaches, XGboost~\cite{chen2016xgboost} and LightGBM~\cite{ke2017lightgbm}  use histograms to speed up training. (Their APIs are available in standard python packages~\cite{histogramboost})   
Using histograms can reduce training time dramatically - e.g.,  Xgboost ~\cite{histogramboost} using histograms takes
  69.8s  on a one-million size table with 10 attributes using 700k training samples and achieves accuracy of 0.985, while it takes only 1s while 
   training on the 20k samples picked by Miao's approach~\cite{miao2021efficient} to achieve 0.971 accuracy. Sampling and histogram methods make learning-based methods
   amenable to online processing by dramatically reducing learning costs. If learning-costs can be brought down to make blocking strategies practical in online settings,  QUIP can bring further improvements by reducing 
   redundant imputations. This is specially the case when the models learnt are relatively complex with non-negligible inference times. 
   Figure~\ref{fig:ML} lists  several machine learning approaches  widely used for missing value imputation\footnote{All of these methods have standard Python packages (in \textsf{sklearn} or \textsf{xgboost}).}. The figure  
   shows  the training time versus total inference time by varying the percentage of missing values (called missing ratio (\textit{MR})  for a table  with 1M rows, 10 attributes where 1 column have missing values. For  approaches with expensive inference, such as SVM and KNN (\textsf{sklearn.impute.KNNImputer}), QUIP achieve 5x to 15x speed-up over ImputeDB in real data by reducing the number of tuples imputed in experiment in Section~\ref{sec:evaluation}.  

In summary, 
QUIP is useful for non-blocking imputation approaches and also for blocking approaches when learning time is reduced (via sampling, histograms, etc.) 
so as 
not to dominate the total inference time, 
or when imputation of individual 
missing value itself is expensive (e.g., rule-based approaches). 




\vspace{-0.5em}
\subsection{Analysis-aware Data Cleaning}
The work most relevant to us is ImputeDB~\cite{cambronero2017query} that we have already
 discussed. 
 As mentioned earlier, while  ImputeDB and QUIP overlap in their goals, they offer
 complementary solutions: while 
 ImputeDB develops a novel optimizer that generates a query plan to  minimize joint cost of imputation and query processing,   QUIP takes a query plan, (e.g., the one generated by ImputeDB), and executes in a way so as to eliminate imputations as much as possible while executing the plan. Our experiments show that by steering the cleaning in the context of the ImputeDB query plan QUIP gets 2 to 10 times of improvement. 
Our work is also related to query-time cleaning approaches as in ~\cite{altwaijry2015query,giannakopoulou2020cleaning,altwaijry2013query}. 
  ~\cite{giannakopoulou2020cleaning} explores  repair of denial constraint violations on-demand to integrate data cleaning into the analysis by relaxing query results.  QDA~\cite{altwaijry2013query,altwaijry2015query} develops a \textit{lazy} strategy for entity resolution to reduce entity resolution overhead  during query processing. The method uses a sketch based approach as a filter to eliminate non-matching  tuples to reduce cost of ER during
  queries. The approach  developed is  specific to ER setting with blocks and there is no effective way to use it for missing value imputation. A naive approach to use the strategy of  ~\cite{altwaijry2015query}  in our setting  (as we will show in  Section~\ref{sec:evaluation}) leads to very high overhead. 

\vspace{-0.5em}
\section{QUIP Overview}
\label{sec:problem}

\begin{figure}[tb]
    \centering
    \includegraphics[width=0.9\linewidth]{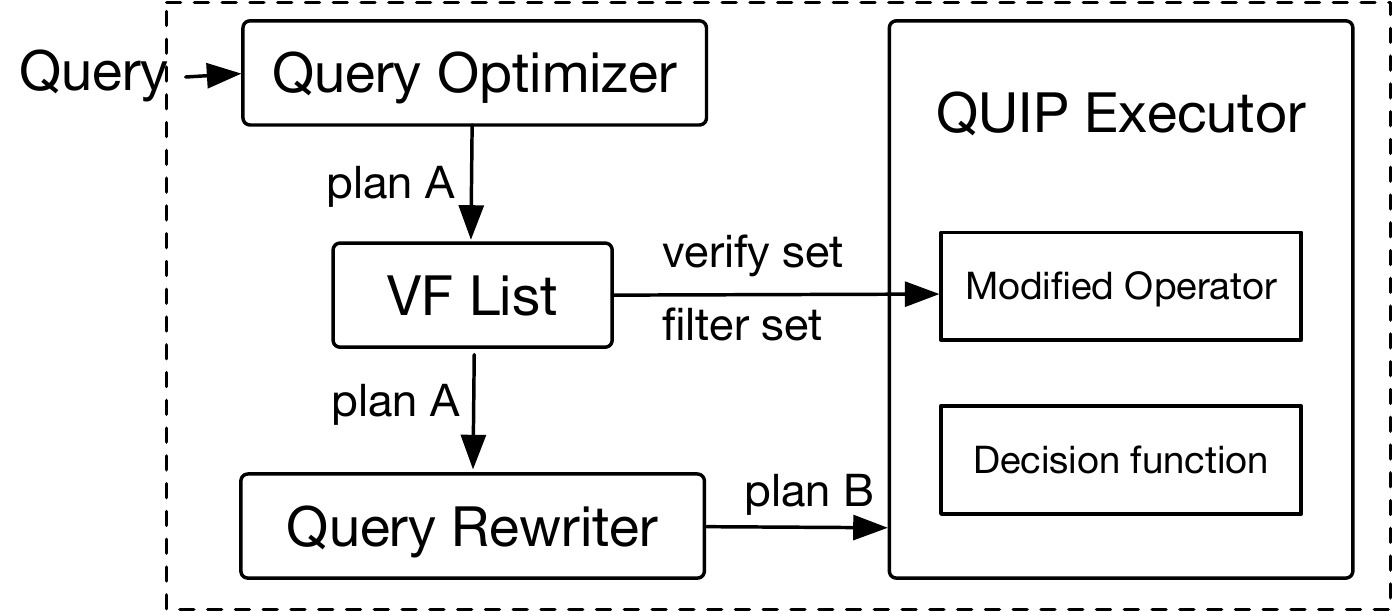}
    \vspace{-1em}
    \caption{QUIP Architecture.}
    \vspace{-1.5em}
    \label{fig:architecture}
\end{figure}

This section provides an overview of   QUIP  that executes lazy evaluation of mixed imputation and query processing. The overall strategy  is 
illustrated in Figure~\ref{fig:architecture}. 

The goal of QUIP is to allow as much laziness as one needs to optimize the joint cost of imputations and query processing, essentially, ensuring \textit{lazy but correct} execution, where correct execution means that the answers returned by QUIP for a query $Q$ is identical to what would be returned had we imputed all the missing values prior to the query execution. 
Consider a query $Q$ (as in Figure~\ref{fig:query}). 
QUIP uses a third-party
optimizer, (e.g., a standard
commercial system such as PostgreSQL or a specialized 
optimizer such as that supported by ImputeDB), 
to first generate
a query plan (plan $A$ in Figure~\ref{fig:architecture}). 
Given a plan,  a data structure entitled VF List is created
that will help guide how 
operators in the 
query plan are modified.
In particular,   for each operator in the plan,
the meta-data stored in the data structure will help determine which tuples
do not need to be imputed  since they do not meet the query conditions associated with the operators downstream. 
After this step, the query plan is passed to a \textit{Query Rewriter} that rewrites the plan using modified operators that QUIP implements. (plan $B$ in Figure~\ref{fig:architecture}).  Query rewriter does not change the structures of the  query plan but replaces each operator with new modified imputation-aware operators supported in QUIP. These modified operators
 can impute, process missing values, or delay imputations to later query processing based on  a cost-based \textit{decision function}.

\begin{figure}[tb]
\vspace{-0.5em}
    \centering
    \includegraphics[width=0.7\linewidth]{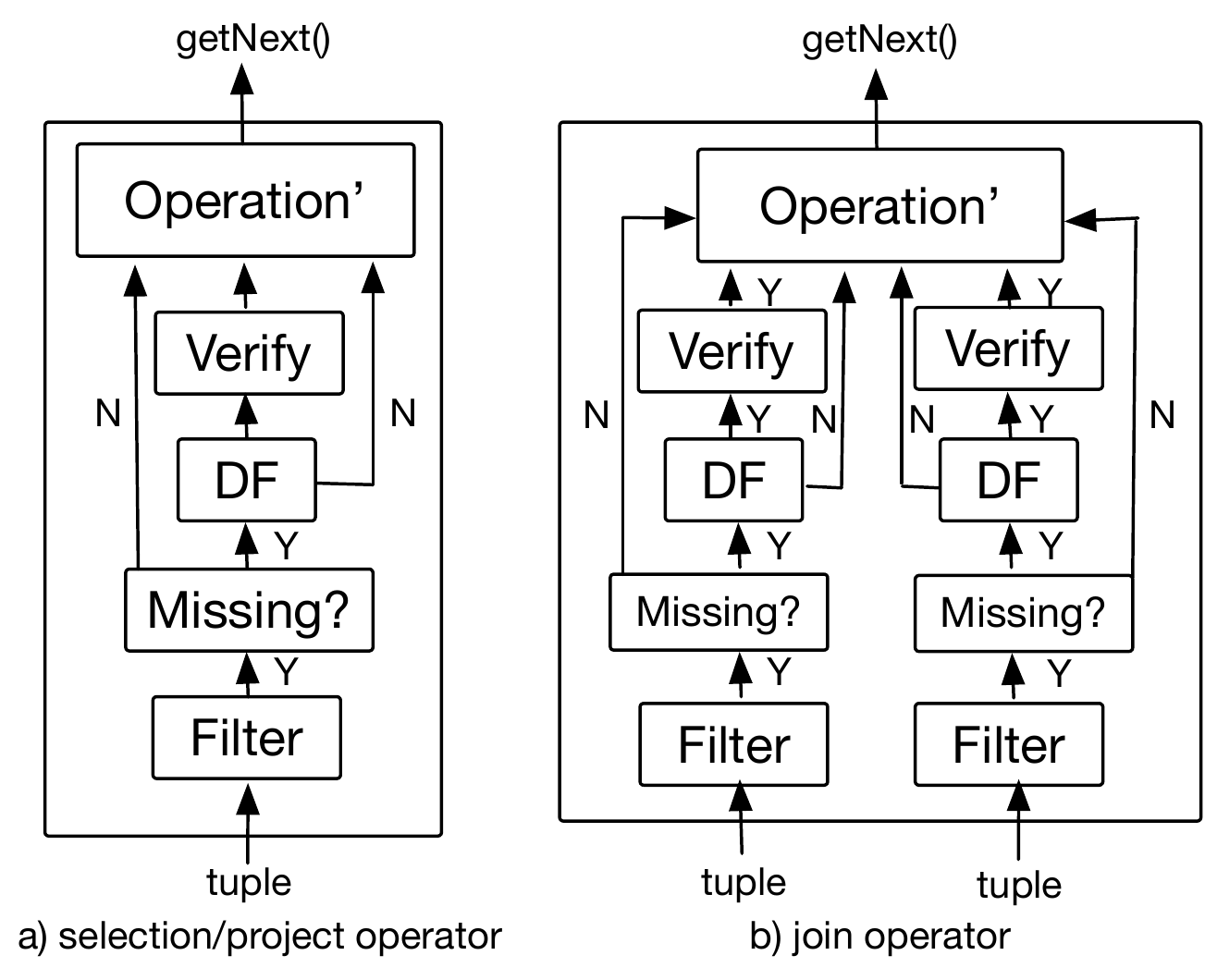}
    \vspace{-1.5em}
    \caption{Modified Operators in QUIP.}
    \vspace{-2em}
    \label{fig:operator}
\end{figure} 

The  flow of
data in the modified selection and join operators are illustrated in Figure~\ref{fig:operator}. 
What QUIP modifies is how  tuples are routed through the operator. 
For each  tuple that passes through an operator, 
the modified operator  first tests a \textsf{filter} condition to  check if this tuple can be filtered  due to an \textit{active} predicate. Such a predicate   helps  determine if there exists a condition associated with a downstream operator that will not be satisfied by the tuple. 
For the tuple $t$ that  passes the filter, if there are no missing values on the attribute $a$ on which the operator is defined,  $t$ is passed to the modified 
operation for processing. A modified operator,  \textsf{{\it operation$^{'}$}}, follows the  underlying implementations of the original operator but makes  minor changes to how it handles NULL values (which may represent missing values). 
Otherwise, if  attribute $a$ on which the operator is defined has a missing value,
a decision function, \textsf{DF} is invoked  to decide whether or to \textit{impute} or \textit{delay} imputing the missing attribute value $v$  prior to executing the operator.  If \textsf{DF} decides to delay the imputation, the tuple is forwarded to the operator. Else, if \textsf{DF} decides to impute $v$, a \textit{verify} step is further performed to check
 if the imputed value $v$ would have failed to pass conditions associated with upstream (i.e., previously executed operators) had those operators seen the imputed value. The tuple with the imputed value is passed to the operator if it succeeds the verification. Details of how filter and verification tests are designed and
how the operators are modified 
become clear in Section~\ref{sec:algorithm}.

\section{Data Structures}
\label{sec:structure}

In this section we describe two data structures used in QUIP, \textit{VF list} and \textit{bloom filter}.

\begin{figure}[tb]
	\centering
	\includegraphics[width=0.9\linewidth]{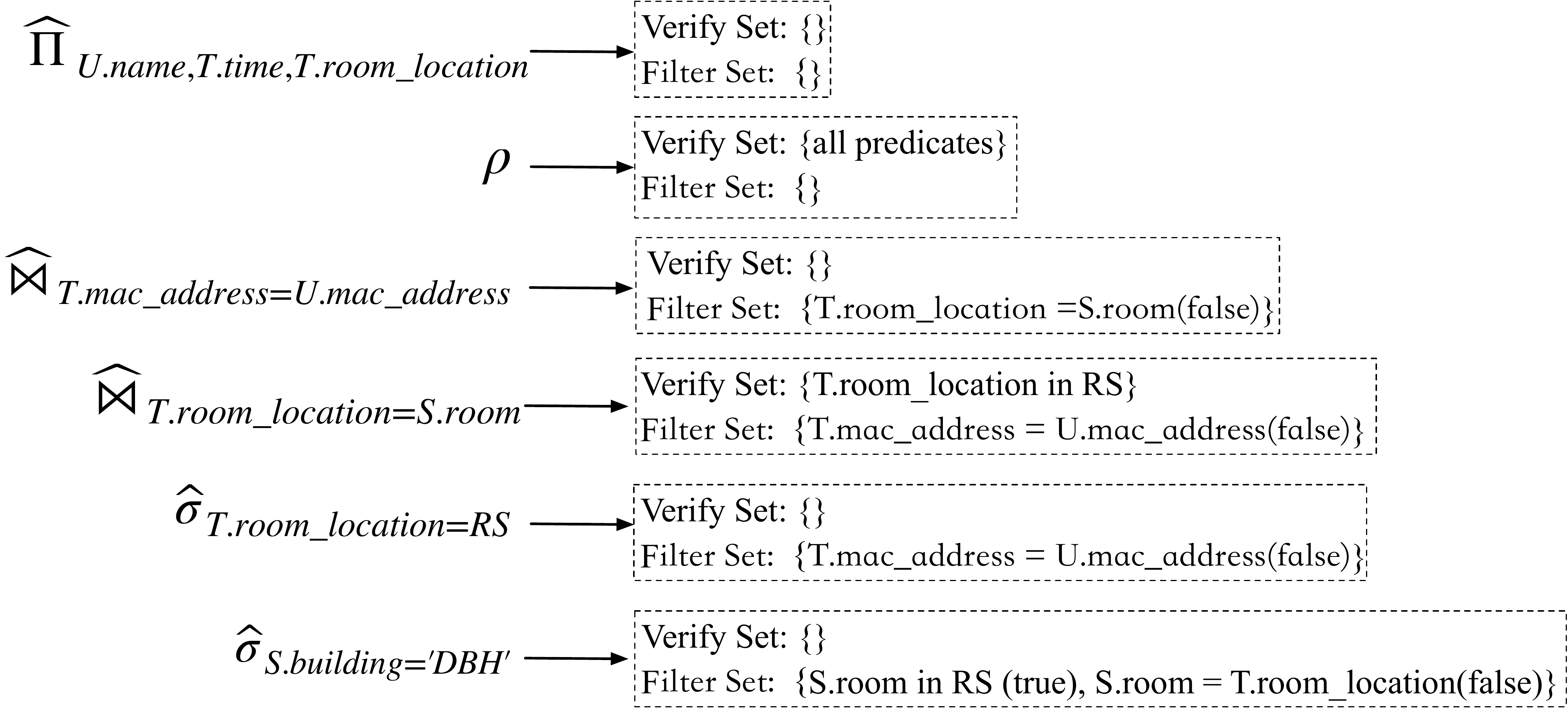} 
	\vspace{-1em}
	\caption{VF List.}
	\vspace{-2em}
	\label{fig:vf}
\end{figure}

\begin{figure*}[tb]
    \centering
    \includegraphics[width=0.93\linewidth]{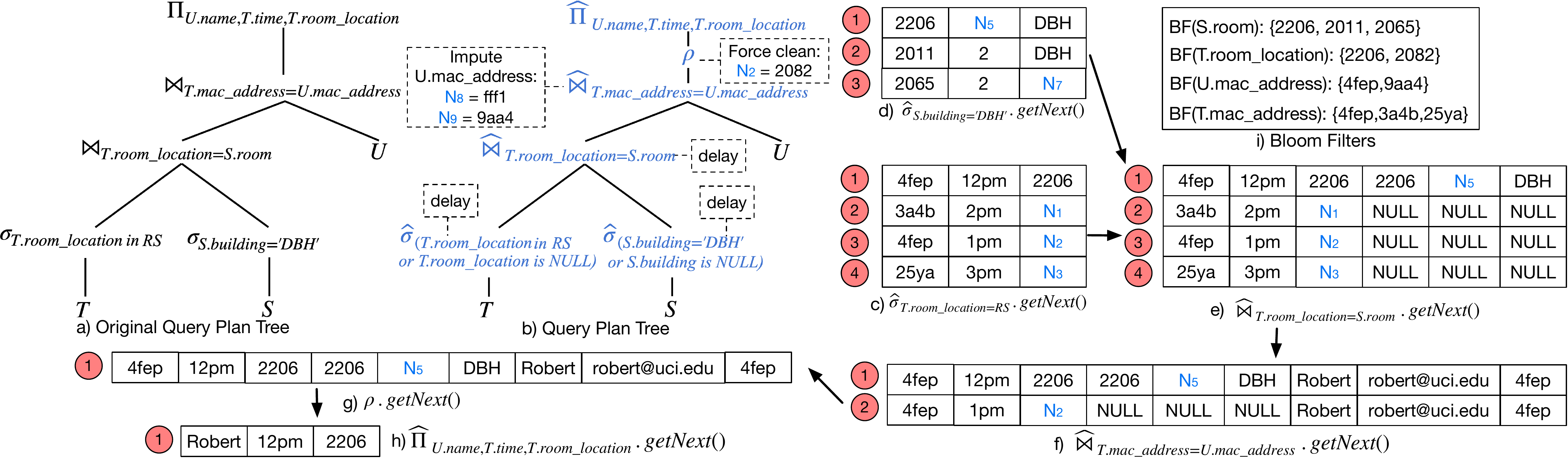}
    \vspace{-1em}
    \caption{Pipeline QUIP.}
    \vspace{-1.5em}
    \label{fig:pipeline}
\end{figure*}

\noindent\textbf{VF List:}
QUIP maintains two sets of 
predicates in the form of \textit{VF List} (see Figure~\ref{fig:vf}) that consist of 
a \textit{verify set} and a \textit{filter set} for each operator in the  query tree (Figure~\ref{fig:pipeline}-a,b is  query tree for query in Figure~\ref{fig:query}.) These 
sets help implement the verify and filter operations respectively 
(see Figure~\ref{fig:operator}). The idea of verify set is simple: whenever a missing value is imputed, it must satisfy the predicates associated with  all previous operators 
that are applicable to it which it  had skipped previously. Let an operator be $o$ (e.g., $\widehat{\bowtie}_{T.room\_location=S.room}$ in the query plan tree (Figure~\ref{fig:pipeline}-a) of the query (Figure~\ref{fig:query})), and the value being imputed correspond to attributes  $A_o$ (e.g., $T.room\_location$ and $S.Room$). 
We construct verify set for each operator $o$ by using all the predicates that are below operator $o$ in query tree and are applicable in attributes $A_o$. For example, the verify set of join operator $\widehat{\bowtie}_{T.room\_location=S.Room}$ is \{$T.room\_location$ $in$ $RS$\}~\footnote{$RS=\{ {2065, 2011, 2082, 2035, 2206}\}$ in query Figure~\ref{fig:query} for short.} since it is applicable to $T.room\_location$ and its associated operator is below the join operator while no predicates are applicable to $S.Room$.  
When a missing value $v$ in $T.room\_location$ is imputed in  $\widehat{\bowtie}_{T.room\_location=S.room}$ operator, QUIP needs to ensure that $v$ must satisfy $T.room\_location$ $in$ $RS$. In Figure~\ref{fig:vf}, $\rho$ is new operator added by QUIP that is put just \textit{above} any selection and join operator in query tree, which will be detailed in Section~\ref{sec:algorithm}. $\rho$ will impute all the missing values in the attributes in query predicates for the received tuple, so its verify set carries all predicates in the upstream operators. 


Filter set, on the other hand, maintains information to enable the \textit{filter} operation  (in Figure~\ref{fig:operator})  to test if a  tuple  $t$ that is input to the (relational) operator $o$ can be filtered away or not.  Let $A_o$ be the attributes associated with $o$. Filter set is created for each operator in the given query plan tree. For each operator $o$, we first find all the predicates associated with its downstream operators that are applicable to the attributes of $t$ \textit{other than} $A_o$. As an example, consider selection operator  $o = \sigma_{S.Building='DBH'}$ in plan tree in Figure~\ref{fig:pipeline}-a) and $A_o = \{S.Building\}$, we first add the predicate  $\{S.Room=T.room\_location\}$ into its filter set because it is from downstream operators of $o$ and it is applicable in relation $S$ other than $S.Building$. Then we extend filter set by finding the \textit{transitive closure} of current filter set. In this example, predicate $\{S.Room$ $in$ $RS\}$ will be added. 
We next associate with each \textit{join predicate} in the filter set a bit (to denote its status) which is initialized to 0. The bit is modified during query processing to indicate when one of the attributes in the join has no remaining missing values. In such a case, the join condition can be used for pruning imputations. We will later develop an efficient mechanism in Section~\ref{subsec:trigger} to use it for filtering. To illustrate how filter operation works, consider the filter set for the above selection operator $\sigma_{S.Building='DBH'}$, i.e., $\{S.Room=T.room\_location, S.Room$ $in$ $RS\}$. 
We can use the selection predicate $\{S.Room$ $in$ $RS\}$ right away for each tuple received by  $\sigma_{S.Building='DBH'}$ to check if this tuple can be filtered first before imputing missing values under $S.Building$ if any. However,  $\{S.Room=T.room\_location\}$ cannot not be used until there are no missing values in $T.room\_location$ during pipeling processing. 

\noindent\textbf{Bloom Filter:} 
QUIP maintains a bloom filter~\cite{bloomfilter} to store attribute values for equi-join operators. 
Let $BF(a)$ be the bloom filter built for attribute $a$. 
In a pipeline implementation, when a tuple rises to the join operator (from either of the two operands), the associated value of the join attribute is inserted into the corresponding bloom filter, if it is
not missing. Also, whenever a missing value for a join attribute is imputed (either as part of the join or a further downstream operator), it will be first verified and added into the corresponding bloom filter if it passes the verification.

We next define a concept of \textit{bloom Filter completeness} with respect to a query $Q$. Intuitively, a bloom filter of an attribute $a$, $BF(a)$, is complete wrt $Q$ if $BF(a)$ contains all the values of $a$ that could \textit{result in tuples in the answer set of $Q$}. 
Note that tuples that are filtered away by the selection/join operators will not appear in the query answer, and hence may not be in the bloom filter.
Formally, we define the completeness of a bloom filter wrt $Q$ as follows. Let $Ans_Q$ the set of tuples that satisfies all predicates of $Q$. 
Note that if $Q$ contains a final projection, some of the tuples in $Ans_Q$ might not appear in the final answer to $Q$. However, for 
the bloom filter $BF(a)$ for 
attribute $a$ to be complete,   values of $a$ in tuples that satisfy all predicates of $Q$  must be in $BF(a)$ even though the tuple is eliminated in the final answer. 
Tuple \circled{1} in  Figure~\ref{fig:pipeline}-g) is the answer set of query $Q$ in Figure~\ref{fig:query}. Consider attribute $S.room$ and $2206$ is its only one attribute value in query answer. 
The bloom filter of $S.room$ in Figure~\ref{fig:pipeline} is complete because it contains $2206$. Note that a complete bloom filter may contain the values that are not query answer, but it must not miss a value that will be part of query answer.

Let $BFC(a)$ be the event that causes the bloom filter $BF(a)$ to be complete. Such an event depends upon the specific implementation of the join algorithm.
Consider a join $L.a\bowtie R.b$, if this join is implemented using 
nested loop, for inner relation $R$, bloom filter $BF(R.b)$ contains all values in $R.b$ 
(i.e., $BFC(R.b)$ is reached) 
when the first pass of relation $R$ is processed. For outer relation $L$, such a condition becomes true only when all 
tuples  have been processed through the join operator. Prior to that we cannot be sure  that
all of $L$ values 
that could result in output from the join are  in the bloom filter $BF(L.a)$. 
For hash joins, similar to nested loop, 
the bloom filter contains all values as soon 
as the hash table based on inner (build) relation
has been built and for outer relation such a condition is reached when all tuples have been processed. For sort merge, or multi-pass hash join, the bloom filters for both relations $L$ and $R$ contains all values as soon as the sort or hash table build is finished. Note that a complete bloom filter $BF(a)$ can only be reached after all the missing values under attribute $a$ have either be imputed or eliminated.

Besides VF list and bloom filter, we also maintain an array, called \textit{missing counter}, that records the number of missing values for each attribute. Such array could be easily initialized using the metadata or statistics in database. Whenever a missing value is imputed or dropped, we will update its corresponding entry in array.  

\section{Imputation and Query Processing}
\label{sec:algorithm}

In this section, we describe a general framework for query processing with imputations in QUIP treating the  decision function in the modified operator as an oracle. Decision functions will be discussed in the next section.  The cost-based design of decision function  relies on estimating cost of
imputing which relates back to how QUIP imputes missing value and executes query.  The strategies to handle blocking or non-blocking imputations in QUIP are controlled by decision function and will also be discussed in Section~\ref{sec:decide}. 
We focus this section on SPJ (select-join-project) query,
discuss  extensions to aggregate operator, union, set difference operators and  nested queries  in Section~\ref{subsec:extensions}. Such extensions are relatively straightforward and 
not discussed in the paper due to length constraints.

QUIP modifies the original query tree input to it (Figure~\ref{fig:pipeline}-a) based on an external optimizer replacing each operator by its modified counterpart without changing the structure of tree. Further, it adds a new imputation operator (denoted by $\rho$) just above any join and selection operation (Figure~\ref{fig:pipeline}-b), which will impute all the missing values in the attributes that appear in the predicates of the given query. SPJ operators are modified by changing how data flows through them as shown in  Figure~\ref{fig:operator}.  

SPJ operators will internally check if a value is missing or not. In QUIP 
missing values are represented as NULLs. We extend the relational schema with an additional attribute which contains a bit per
attribute of the relation. If the value of  attribute $a$ in a tuple $t$ is NULL, and the corresponding bit is set to 1, we treat that NULL value of $a$ as missing, else (i.e., the bit is not set) it is treated as a regular NULL. Using the bit by suitably modifying the associated predicate, the operators 
can differentiate  missing values from NULLs. 
Henceforth (e.g., in Figure~\ref{fig:pipeline})  we will also differentiate missing and NULL values. 

Operators in QUIP are  implemented using an \textit{Iterator Interface}. 

\noindent$\bullet$ open(): initializes the operation. \\
\noindent$\bullet$ getNext(): return the next tuple that is satisfied by the modified operators to its father without materializing it. \\
\noindent$\bullet$ close(): all tuples are produced and close the iterations. 


\vspace{-0.7em}
\subsection{Modified SP Operators}
\noindent\textbf{Modified Selection Operator}, denoted by $\widehat{\sigma}$. Consider a selection operation $\widehat{\sigma}_{pred}$, where $pred = a$ op $v$, where  $a$ is an attribute and $v$ is a value, and let  $t$ be the tuple 
that has ascended to this operator. If $t$ passes the \textit{filter} test and $t.a$ is not missing (i.e., the value is not NULL), then $\widehat{\sigma}_{pred}$ simply checks if $t.a$ satisfies the predicate $pred$. 
If $t.a$ is missing and the decision function  decides to impute $t.a$, then the verify set and predicate will be used to check the imputed value. Note that we can check for $t.a$ to be missing by adding an appropriate disjunction in the predicate (as shown in example below). If decision function decides to delay imputation or $t$ passes the verify test and the predicate $pred$, then $t$ will be returned by $\widehat{\sigma}.getNext()$. Otherwise, $t$ will be discarded and $\widehat{\sigma}.getNext()$ will return the next  tuple that satisfies the  $\widehat{\sigma}$. 

\begin{example}
Figure~\ref{fig:pipeline} illustrates the pipeline implementation of query in Figure~\ref{fig:query} in QUIP. In Figure~\ref{fig:pipeline}-c) to h), the numbered red circle represents the order in which $getNext()$ function returns these tuples for each operator.  Figure~\ref{fig:pipeline}-b) states the decisions taken by the decision functions in each operator. In selection operator, such as $\widehat{\sigma}_{S.Building='DBH'}$, we add another condition $S.building$ $is$ $NULL$ to read the missing values from relation $S$. ~\footnote{If index scan is implemented in scan operator, NULL values would be stored in index and be returned. This is supported in most current commercial databases,  such as  PostgreSQL~\cite{postgresql}.} 
First, QUIP decides to delay imputations in two selection operator  $\widehat{\sigma}_{T.room\_location \in RS}$ and $\widehat{\sigma}_{S.Building='DBH'}$, and their \textsf{getNext()} tuples are shown in  Figure~\ref{fig:pipeline}-c) and Figure~\ref{fig:pipeline}-d), respectively. Note that the tuples in Space table (Table~\ref{tab:S}) with Room value 2214 and 3119 are dropped because they failed to pass the filter set of  $\widehat{\sigma}_{S.Building='DBH'}$ in Figure~\ref{fig:vf}, i.e., $\{S.Room$ $in$ $RS = \{2065,2011,2082,2035,2206\}\}$. 
\end{example}

\vspace{-2mm}
\noindent\textbf{Modified Projection Operator}, denoted by $\widehat{\Pi}$. To keep it simple, we consider the case where projection operator is at the top of query  tree. In this case, for a tuple $t$ received by $\widehat{\Pi}$, if $t$ passes the filter test using filter set in VF list, then $\widehat{\Pi}$ will force  imputation of all missing values under projected attributes if any, and the verify set will be called to verify the imputed tuple. The tuple $t$ that passes the filter and verify tests will be returned by $\widehat{\Pi}.getNext()$. Otherwise, $\widehat{\Pi}$ will search next satisfied tuple to return until all tuples are consumed. 

\vspace{-0.5em}
\subsection{Modified Join Operator}
\label{subsec:join}

\begin{figure}[tb]
	\centering
	\includegraphics[width=1\linewidth]{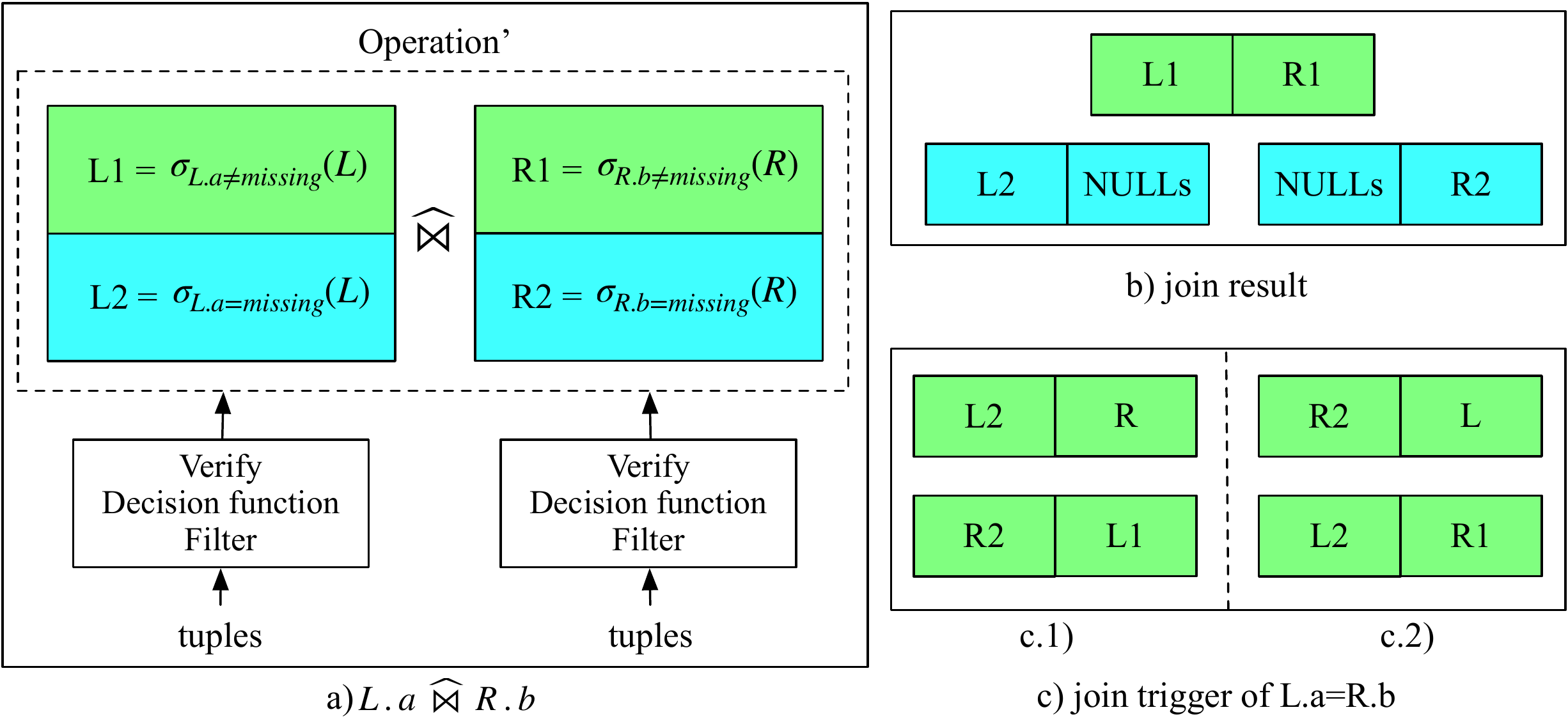}
	\vspace{-2.5em}
	\caption{\textbf{$L.a$ $\widehat{\bowtie}$ $R.b.$}}
	\vspace{-2em}
	\label{fig:join}
\end{figure}

We denote a modified join operator by $\widehat{\bowtie}$. 
 Consider a join between relations 
$R^L$ and $R^R$ based on attributes $a$ and $b$ respectively (i.e.,
$R^{L}.a$ $\widehat{\bowtie}$ $R^{R}.b$), where 
$R^L$ and $R^R$ are the 
left and right relations respectively, and $a$ and $b$ are join attributes.  When there is no ambiguity, we
will refer to $R^L$ and $R^R$ simply as $L$ and $R$.
Consider join operator $L.a\widehat{\bowtie}R.b$, and  a tuple $t$ in either $L$ or $R$. For each such tuple, 
QUIP  executes a \textit{filter}, \textit{decision function} and the \textit{verify} operations as shown in  Figure~\ref{fig:operator}-b). 
Any tuple that fails to pass the filter or verify set will be dropped. We focus our discussions on the modified actions ($operation^{'}$ box in Figure~\ref{fig:operator}-b) for those tuples that pass filter and verify tests. We explain the $operation^{'}$ in Figure~\ref{fig:join}-a). 
Tuples that arrive to $operation^{'}$ pass the filter and verify nodes on either left or right side of the 
input can be classified into two types:

    \noindent1) Tuples that do not contain missing values in the join attribute. (denoted as $L1$ or $R1$ in Figure~\ref{fig:join}-a)\footnote{Tuples in $L1$ (or $R1$) might have contained missing values for the join attribute prior to reaching $operation^{'}$, but for such tuples, those missing values  must have been imputed and passed the verify test (else, 
such tuples will not be of type 1). } 
     2) Tuples that contain missing values in the join attribute. (denoted as $L2$ or $R2$ in Figure~\ref{fig:join}-a)

Consider join operation $L.a \bowtie R.b$, as shown in Figure~\ref{fig:join}-a), $L.a\bowtie R.b$ can be rewritten as $L1 \bowtie R1$ $\cup$ $L1 \bowtie R2$ $\cup$  $L2\bowtie R1$ $\cup$ $L2 \bowtie R2$. 
We next discuss how  join is implemented
based on whether  neither, one or
both  attributes in the join contain 
missing values. 
If there are missing values in both or one side join attribute, then a full/left(right) outer join based modification is implemented for join algorithm as not to lose any missing value. If neither join attributes have 
missing values, then regular join with one
difference that if there are tuples from previous outer-join which contain NULL value in join attribute, we need to pass this tuple above.

\noindent {\bf Case 1: Both side $L.a$ and $R.b$ do not contain missing values. ($L2=\emptyset$ and $R2=\emptyset$)}. In this case, $L.a\bowtie R.b = L1 \bowtie R1$, and tuples will be joined using the original join implementation in the operator. If these tuples are the outer-join result of previous tuples (they contain NULL values under join attributes), then QUIP simply lets them pass by concatenating NULL values for the attributes corresponding to  the other side of the join operation. Such tuples arise if, for instance, 
a previous join operator had decided not to 
impute a missing value in its join attribute for this tuple. 
Otherwise, they will be joined as the original join condition. 
Consider join operation  $\widehat{\bowtie}_{T.mac\_address=U.mac\_address}$ in Figure~\ref{fig:pipeline}-b), assuming the decision function decides to impute missing values $N_8$ and $N_9$ in attribute $U.mac
\_address$, tuples in $T$ relation in  Figure~\ref{fig:pipeline}-e) will join relation $U$ in Table~\ref{tab:U}, which leads to  Figure~\ref{fig:pipeline}-f). 

\noindent {\bf  Case 2: Only one side attribute in join operation contains missing values. ($L2=\emptyset$ or $R2=\emptyset$)} Without loss of generality, let us consider the case where the left side of join attributes, $L.a$, has missing values. In this case, if tuples on the left side are of type $1)$, they will be joined with right side based on the original join predicate ($L1 \bowtie R1$ $\cup$ $L1 \bowtie R2$). Else, if the tuple in left side is of type $2)$, they will be returned as part of outer join by concatenating NULL values in the right columns. Consider join operator $\widehat{\bowtie}_{T.room\_location=S.room}$ in Figure~\ref{fig:pipeline}-e), where only $T.room\_location$ has missing values. 
The tuple \circled{1}  in Figure~\ref{fig:pipeline}-e), is a joined tuple from tuple \circled{1} in $T$ relation in Figure~\ref{fig:pipeline}-c) and tuple \circled{1} in $S$ relation in Figure~\ref{fig:pipeline}-d). All the other tuples, i.e., tuples \circled{2}-\circled{4} in Figure~\ref{fig:pipeline}-e), are the left join results of $T$ and $S$ where the missing values $N_1,N_2,N_3$ are preserved with NULLs in columns in $S$ side. Note that we will not lose the missing value $N_7$ in tuple \circled{3} in Figure~\ref{fig:pipeline}-d) although it does not show up in later query processing, because it will be brought back if the $S.building$ value in tuple \circled{3}, i.e., 2065, is matched with any attribute value in $T.room\_location$ by using  the bloom filter $BF(T.room\_location)$. In Section~\ref{subsec:trigger}, we will detail how to achieve this using bloom filters. 

\noindent {\bf Case 3: $L.a$ and $R.b$ both 
contain missing values in the join attribute. ($L2 \neq \emptyset$ and $R2 \neq \emptyset$)} In this case, tuples of type $1)$ and $2)$ can be in both sides of the join column. Tuples of type $1)$ will be joined using the original join predicate in the operator ($L1 \bowtie R1$).  Tuples of type $2)$, that is, with missing values, will be preserved as in the outer join. That is, 
 if a tuple of $L$ contains a missing value in $L.a$, the modified operator will generate a tuple with values of $L.a$ and with NULLs in all the attributes corresponding to $R$. Likewise, a similar tuple would be created with NULLs for attributes on $L$ side for all tuples in $R$ that arrives to the operator with a missing value for $R.b$. In this case, QUIP maintains
the list of tuple-id, $\mathcal{L}_{temp}$ for tuples 
in $L2$ or $R2$ whichever has lesser 
number of tuples, and use a $Flag$ to denote the side of relation to be stored (e.g., $Flag = L$ denotes tuple-ids for $L2$ is stored).  The cardinality of 
$L2$ and $R2$ is simply the number of missing values in $L.a$ and $R.b$, which corresponds to their respective missing counters. 
The reason for maintaining this set will become clear later when we discuss join triggers.

In realizing the implementation of the modified join operation, note that QUIP
can exploit the existing join implementation(s) supported by
the database in which QUIP is integrated 
since it 
only modifies
how data flows through the operator and not the operator implementation. 
 For instance, if the database implements   a hash join, QUIP  continues to use the hash join implementation for tuples with type $1)$. The only addition it makes is 
 to check if an input tuple to the join 
 has a  NULL value in join attribute, in which case, it simply forward this tuple to the 
 next downstream operator by first suitably
 padding NULLs in the requisite fields like
 the way an outer-join does. 
 


\vspace{-1em}
\subsection{Join Trigger}
\label{subsec:trigger}
Imputation of missing values in the join operation  $L.a\bowtie R.b$ may result in QUIP triggering the execution of parts of the join that had not executed 
prior to the imputation (i.e., $L1 \bowtie R2 $, $L2 \bowtie R1 $, and $L2 \bowtie R2 $). 
QUIP triggers the execution of such partially 
executed joins  as soon as enough imputations
have been performed to make such a 
join  possible. 
Consider
a join $ L.a \bowtie R.b$, the join that had not executed before, say $L2 \bowtie R1$, might  be triggered when missing values in $L2$ are imputed. 
To compute 
such a join, QUIP exploits the previously constructed bloom filter on $R$ to determine which tuples of $L2$ join. 
Note that since QUIP is based on streaming model, to compute $L2 \bowtie R1$, in addition
to ensuring that all of the tuples of 
$L2$ are imputed (or eliminated), it is also
necessary to ensure that the bloom filter
for $R$ is complete, i.e., $BFC(R.b)$ is true, as defined in Section~\ref{sec:structure}.

We describe join trigger of the join operation $L.a\bowtie R.b$ in Algorithm~\ref{alg:trigger} in format of ECA rules~\cite{mccarthy1989architecture,tan1999implementing}, i.e., 
\textit{when}  \textsf{event}, \textit{if} \textsf{condition}, \textit{then} \textsf{action}. For each tuple $t \in L$ whose $t.a$ value was missing and had not been imputed during the execution of 
$\widehat{\bowtie}_{L.a=R.b}$, when 
$t.a$ is imputed,
if $BFC(R.b)$ is true, then QUIP will use $BF(R.b)$ for join by calling $BF\_Join(t,BF(R.b),\mathcal{L}_{temp},Flag)$  (Ln.2-3).  


\setlength{\textfloatsep}{0pt}
\begin{algorithm}[bt]
    \small
	\caption{Join Trigger of $L.a\bowtie R.b$.}
	\label{alg:trigger}
	\KwIn{$\mathcal{L}_{temp}, Flag$}  
	\SetKwProg{Fn}{When}{:}{}
	   \Fn{$t.a$ is imputed}{
	        \eIf{$BFC(R.b)$ is true}{
	            $BF\_Join$($t, BF(R.b), \mathcal{L}_{temp}, Flag$)
	        }{
	            $\mathcal{L}_{L\_Ready}  \leftarrow \mathcal{L}_{L\_Ready} \cup $ $\{$tid of $t\}$
	        }
	   }
	   \Fn{$t.b$ is imputed}{
	        \eIf{$BFC(L.a)$ is true}{
	            $BF\_Join$($t, BF(L.a), \mathcal{L}_{temp}, Flag$)
	        }{$\mathcal{L}_{R\_Ready}  \leftarrow \mathcal{L}_{R\_Ready} \cup$ $\{$tid of $t\}$
	        }
	   }
	   \Fn{$BFC(L.a)$ is true}{
	        \For{$t\in \mathcal{L}_{R\_Ready}$}{
	            $BF\_Join$($t, BF(R.b), \mathcal{L}_{temp}, Flag$)
	        }
	        $\mathcal{L}_{R\_Ready} \leftarrow \emptyset$
	   }
	   \Fn{$BFC(R.b)$ is true}{
	        \For{$t\in \mathcal{L}_{L\_Ready}$}{
	            $BF\_Join$($t, BF(L.a), \mathcal{L}_{temp}, Flag$)
	        }
	        $\mathcal{L}_{L\_Ready} \leftarrow \emptyset$
	   }

\end{algorithm}

In $BF\_Join(t,BF(R.b),\mathcal{L}_{temp},Flag)$ in Algorithm~\ref{alg:bfjoin}, $t$ will be dropped if $BF(R.b)$ returns false. Note that if bloom filter returns false, $t$ will not be joined with any tuple in relation $R$. 
Otherwise, $BL(R.b)$ will return matched tuples, called $T_{matched}$, in relation $R$ by executing index look ups. (Ln.4) ~\footnote{Index is assumed to be built for join attributes, otherwise QUIP can alternatively  maintain additional linear pointers from attribute values to their tuples.} Next the tuples with tuple-ids $\mathcal{L}_{temp}$ will be removed from  $T_{matched}$ if $\mathcal{L}_{temp}$ and tuple $t$ are in different relations of the triggered join operation. (Ln.5-6) \footnote{Simply doing join of $t\in R2$ using $BF(L.a)$ and $t\in L2$ using $BF(R.b)$ will generate duplicated join tuples since the tuples from $L2\bowtie R2$ will be produced twice. To remove such duplicated results, we use $\mathcal{L}_{temp}$, the stored tuple-ids in the relation with smaller number of missing values, say $R2$, to remove joined tuples originated from $R2$ to implement $L2\bowtie R1$. On the other hand, tuples $t\in R2$ is joined using $BF(L.a)$, i.e., $R2\bowtie L$, which makes the join $L.a\bowtie R.b$ complete.}  

When $t.a$ is imputed, if $BFC(R.b)$ is false, then QUIP will insert the tuple id of $t$ into a list, denoted as $\mathcal{L}_{L\_Ready}$, such that such tuples will be joined later when $BFC(R.b)$ becomes true. (Ln.4-5 in Algorithm~\ref{alg:trigger}.) 
Similarly, when $t.b$ is imputed, if $BFC(L.a)$ is true, then QUIP will use the complete bloom filter $BL(L.a)$ to join tuple $t$ with relation $L$ using $BF\_Join$($t, BF(L), \mathcal{L}_{temp}, Flag$ (Ln.7-8). Else, if $BFC(R.b)$ is false, then QUIP will insert the tuple id of $t$ into a list, denoted as $\mathcal{L}_{R\_Ready}$ (Ln.9-10).  

When $BFC(R.b)$ (or $BFC(L.a)$) becomes true, (Please refer to  Section~\ref{sec:structure} where we describe how to detect when $BFC()$ is true) QUIP performs join for all tuples in $\mathcal{L}_{L\_Ready}$ (or $\mathcal{L}_{R\_Ready}$) one by one calling $BF\_Join(.)$, and $\mathcal{L}_{L\_Ready}$ (or $\mathcal{L}_{R\_Ready}$) will be set as empty to make sure such join is executed only once (Ln 11-18).

\begin{example}
Consider join operation $T.room\_location = S.room$, which is delayed due to delaying imputations in attribute \\$T.room\_location$, and only one side of attribute $T.room\_location$ has missing values. Assume hash join is implemented in \\$\widehat{\bowtie}_{T.room\_location = S.room}$ and $S.room$ is the build relation. In this case, the bloom filter $BF(S.room)$ is complete, i.e., $BFC(S.room)$ is true, after the hash table is built. For every tuple $t$ received by the downstream operators of $\widehat{\bowtie}_{T.room\_location = S.room}$, when the missing value under $T.room\_location$ is imputed, e.g.,  $N_2$ in tuple \circled{2} in Figure~\ref{fig:pipeline}-f) (being imputed in $\rho$ to be 2082 in Figure~\ref{fig:pipeline}-b), QUIP uses $BF(S.room) = \{2206,2011,2065\}$ to execute join for tuple \circled{2}. Tuple \circled{2} will be dropped since its  $T.room\_location$ value is 2082 which does not match any values in $BF(S.room)$, and thus no matched tuples are found. Tuple \circled{1} in Figure~\ref{fig:pipeline}-f will be preserved as it is the joined tuple in $\widehat{\bowtie}_{T.room\_location = S.room}$ operator. 
\end{example}

\setlength{\textfloatsep}{0pt}
\begin{algorithm}[bt]
    \small
	\caption{$BF\_Join(t,BF(a),\mathcal{L}_{temp},Flag)$}
	\label{alg:bfjoin}
	\eIf{$BF(a) = \emptyset$}{
	    $t$ is eliminated;
	}{
	$T_{matched} \leftarrow Index\_look\_up(BF(a))$\\
	\If{($t\in R^{R}$ and $Flag = L$) or ($t\in R^{L}$ and $Flag = R$)}{$T_{matched} \leftarrow T_{matched}\setminus \mathcal{L}_{temp}$ }
	\textbf{return} $t\bowtie T_{matched}$
	}
\end{algorithm}

\vspace{-0.5em}
\noindent\textbf{VF list Update}. 
When $BFC(L.a)$ ($BFC(R.b)$) is true, such a complete bloom filter will not only be used to re-execute join, but also help drop temporary tuples and potentially reduce imputations to speed up query processing. When attribute $R.b$, such as $T.room\_location$, is activated, the VF list, as shown in Figure~\ref{fig:vf}, will set the status of its associated predicate $S.room=T.room\_location$ to be \textit{true}. 
For any operator $o$ that contains $S.room=T.room\_location$ in its filter set, such as $\widehat{\sigma}_{S.building='DBH'}$, QUIP uses bloom filter $BF(T.room\_location)$ to check if $t$ can be filtered away as follows. For tuple $t$ received by $\widehat{\sigma}_{S.building='DBH'}$, if $S.room$ is not missing and not NULL, we use $BF(T.room\_location)$ to check if $S.room$ has any matched values in $BF(T.room\_location)$. If $BF(T.room\_location)$ returns false, we drop tuple $t$. Else, we \textit{do nothing} and let $t$ pass. If $S.room$ is missing, we call decision function and forward tuple $t$ to later operations. Note that, in filter test, we only use bloom filter to do one-side checking. This check operation is safe because bloom filter does not have false negative, and it is also cheap (constant time). We do nothing when matches are found in bloom filter because there is no reason to do join in advance in selection operator $\widehat{\sigma}_{S.building='DBH'}$, which is also not feasible. Such tuple $t$ will be evaluated by join condition $S.room=T.room\_location$ anyway because it comes from the downstream operator.


\noindent\textbf{Discussion}: 
When a tuple with multiple missing values in attributes involved in query predicates reaches imputation operator $\rho$, we first impute and evaluate all the missing values whose attributes are in \textit{selection predicates}. \footnote{This can be determined by looking up their verify set in VF list.}  If all of such imputed values passed selection predicates, then we impute and evaluate missing values involved in join predicates. 
Imputing attribute values in the selection predicates helps reduce query processing cost. The selection predicates are more selective than the join predicates.  
Alternative strategy can be first imputing the missing value with lowest imputation cost, which can be easily estimated as the average imputation cost of missing values under their attribute during query processing. 
\section{Decision Node}
\label{sec:decide}

Recall that in QUIP, each modified relational operator includes a decision node to help decide whether a missing attribute value should be imputed prior to the execution of the operator or should imputation be delayed for downstream operators. 
 The strategy used to decide depends upon whether the imputation method is non-blocking or blocking. 

We focus on an \textit{adaptive} cost-based solution for non-blocking imputations, and  describe \textit{eager} and \textit{lazy} strategies for blocking imputations for short. 
Eager strategy imputes missing values of the attribute(s) associated with the operator right away prior to evaluation of the operator, while lazy strategy always delays imputing until the tuple with the missing value reaches  the imputation operator $\rho$. 
These two strategies are discussed and detailed in Section~\ref{subsec:blocking}.

\vspace{-0.5em}
\subsection{Obligated Attributes}
Non-blocking imputations in QUIP can be placed anywhere in the query tree
since QUIP, through operator modification, decouples imputation from the operator implementation.  To guide the actions of each operator, we
first define a concept of 
{\emph obligated} attributes for relations in query $Q$.  Intuitively, an attribute $a$ in $R$ is obligated if missing values of $a$ in  $R$ "must" be imputed in order to answer the query, i.e., for such attributes its values  cannot be 
eliminated as a result of other  query conditions or due to imputation of other missing values.


\vspace{-1mm}
\begin{definition}
\label{theo:dead}
{\bf (Obligated Attributes)}
Given the set of attributes in predicate set of a query $Q$ (denoted by $A_Q$), an attribute $a$ in relation $R$ is said to be obligated if 
\begin{itemize}
\item attribute $a$  appears in a predicate in $Q$, i.e., $a\in A_{Q}$, or $a$ is one of the  attributes listed in a projection operator;  and
\item all attributes of $R$ (other than attribute $a$) do not appear in any predicate in 
$Q$.  
That is,   $\forall a^{'}\in R-a$,
$a^{'}\notin A_Q$. 
\end{itemize}
\end{definition}
\vspace{-1mm}

If an attribute $a \in R$ is neither in the projection list  nor in $A_Q$, imputing its missing values will not be required to answer $Q$ and hence $a$ would not be obligated. Likewise, if a predicate in $A_Q$ contains
an attribute $b$ which is also in $R$, it is possible that such a predicate may result in the tuple of $R$ to be eliminated thereby making imputation of the corresponding $a$ value (in case it was missing) 
unnecessary. Thus, again, such a possibility would prevent  $a$ from being classified as obligated. 
As an example in Table~\ref{tab:U}, \textsf{U.mac$\_$address} is a obligated attribute because other attributes \textsf{U.name} and \textsf{U.email} are not in any predicate of query $Q$ and \textsf{U.mac$\_$address} is in join predicate \textsf{U.mac$\_$address=T.mac$\_$address}.

Since missing values of obligated attributes must always be imputed,  there is
 no benefit in delaying their imputations. In contrast, imputing could  potentially reduce
number of tuples during query processing. As a result, decision function in the  
QUIP operators never  delay  such imputations. 
For the remaining attributes, QUIP performs a cost-benefit analysis 
to decide whether to impute. 

\vspace{-0.5em}
\subsection{Decision function}
For each operator $o$ in query tree, QUIP associates a decision function $df(a,o)$ for all attribute $a$ that appears in the predicate associated with $o$. 
Decision to delay/impute missing values has implications on both imputation  and query processing costs. 

\begin{figure}[tb]
    \centering
    \includegraphics[width=0.95\linewidth]{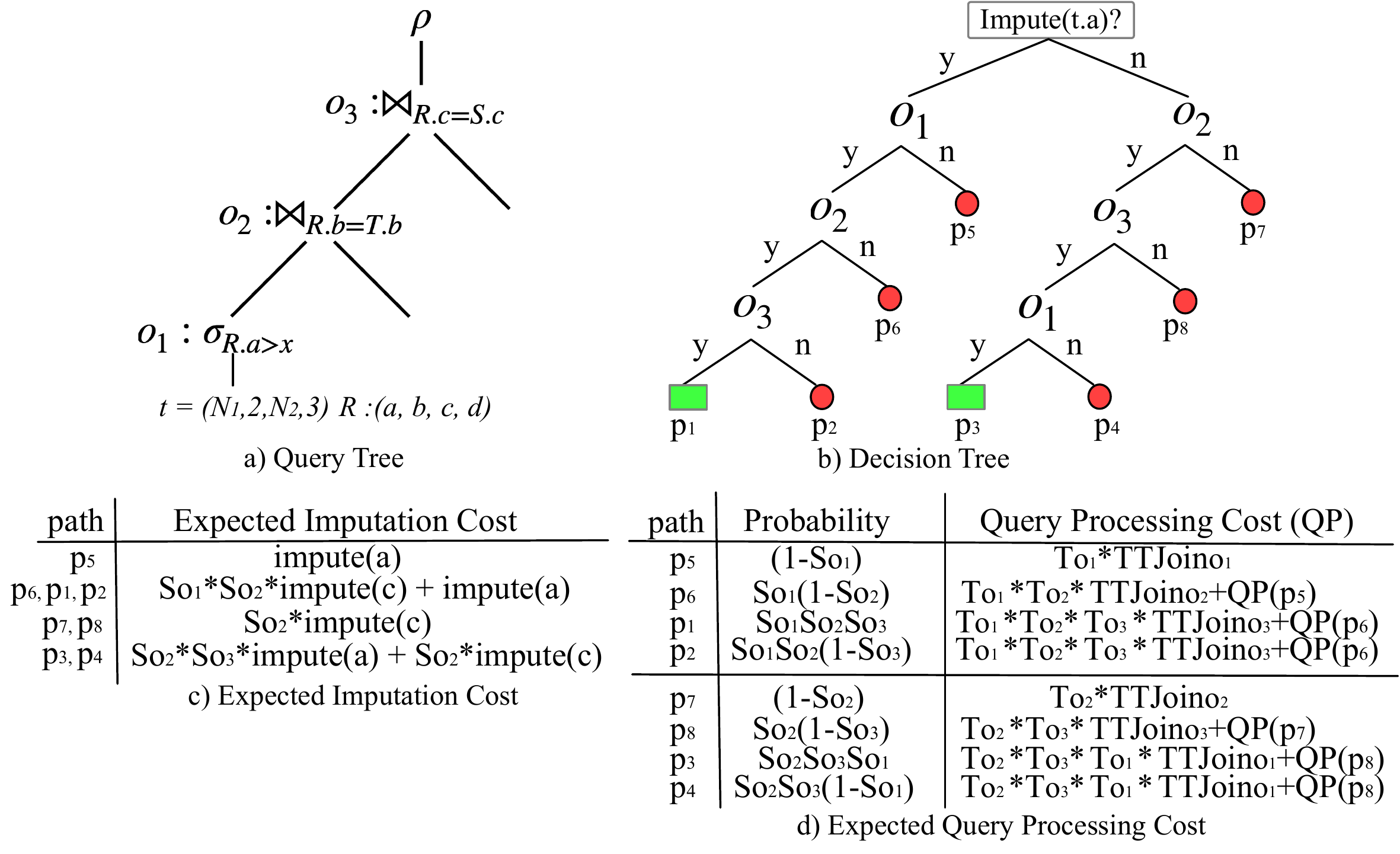}
    \vspace{-1.5em}
    \caption{\small Decision Function Example.}
    \label{fig:decision}
\end{figure}

Consider a tuple $t$ in relation $R=(a,b,c,d)$ and a query tree in Figure~\ref{fig:decision}-a). Say $t_1 = (N_1,1,2,3)$ ($N$ represents missing value), if we delay imputing  $t_1.a$, and $t_1.b$ does not join with any tuples in the other relation, we can avoid imputing $t_1.a$. On the other hand, imputing $t_2.a$ for  $t_2=(N_1,N_2,2,3)$, could prevent imputation of $t_2.b$, if the imputed value of $t_2.a$ is filtered  in the selection operator. 
Imputing $t_2.a$ may also reduce 
query processing time since it does 
not require the operator on attribute $b$ to be executed. 

Since decisions on whether to impute/delay  are made per tuple containing missing values locally by the operator, the decision function  must not incur significant overhead. In making a decision for  operator $o_1$ over attribute value
$t.a$  of a tuple $t$, QUIP assumes if $t$ contains other missing values in attributes, say $t.b$ (on which predicates are defined 
in downstream operators), say $o_2$, those operators will decide to 
impute $t.b$ if (and when) the tuple $t$ reaches those operators. For instance, in query tree in Figure~\ref{fig:decision}-a), in making a decision for imputing /delaying  $t.a$, i.e., $N_1$, in developing a cost model we assume that the missing value $N_3$ ($t.c$) will be imputed right away.  This prevents, QUIP to have to recursively consider a larger search space that enumerates (potentially exponential number of other possibilities wherein downstream operators may delay/impute.)

We build a cost model below to estimate impact of delay/impute decision on both the imputation cost and the query processing cost based on which the operators make  decisions in QUIP.
To compute the imputation and query processing 
costs associated of the decision for an operator, QUIP maintains the following statistics: 

    \noindent
    $\bullet$ $impute(a)$: Cost of imputing a missing value of attribute $a$, 
      computed as a running average over all imputations performed so far for missing values of $a$.  \\
    \noindent
    $\bullet$ Selectivity of selection operator $o_i$, $S_{\o_i} = \frac{|T_s|}{|T_c|}$, where $T_c$
    ($T_s$) are tuples that are processed (satisfy) the predicate associated with $\o_i$. \\
    $\bullet$ Selectivity of join operator between relation $L$ and $R$ computed as  $S_{o_i} = \frac{|T_s|}{|T_L||T_R|}$, where $T_L$ ($T_R$) are tuples in relation $L$ ($R$) and $T_s$ are tuples that satisfy $o_i$ \footnote{We exclude those tuples containing missing  values from $T_s$, $T_c$, $T_L$ and $T_R$.} \\ 
    \noindent
   $\bullet$ $TTJoin_o$: the average time taken to join two tuples in (join) operator $o$; \footnote{We also use the notation $TTjoin_o$ for  selection operator, in this case,  $TTjoin_o=0$.} \\
    \noindent
    $\bullet$ $\mathcal{T}_o$: the average number of evaluation  tests to perform per tuple in operator $o$ for tuples without missing values in the attribute to be evaluate in $o$.~\footnote{A tuple with missing value in the attribute that passes through $o$ will be simply preserved to the above operator without evaluation immediately, and thus $\mathcal{T}_o$ for such tuple is set to be 1.}  If $o$ is join operator, evaluation tests refer to join tests. Else, if $o$ is selection operator, we set  $\mathcal{T}_o=1$. 

 To bootstrap the process of statistics collection, QUIP initially delays all imputations forcing tuples to rise up to the top of the tree (or be dropped if
 they fail some predicates en-route).  
 During this process, QUIP  collects imputed tuple samples to compute $impute(a)$ and to determine other statistics such as $\mathcal{T}(o)$, join cost $TTJoin_o$ and selectivity $S_{o_i}$. These statistics are then 
 adaptively updated during query processing.

\noindent\textbf{Cost Model for Imputations.} 
We illustrate how to estimate the imputation cost using an example, and include the mathematical  model for imputations and query processing (below) in Section~\ref{subsec:costmodel}. 
Consider a query tree in  Figure~\ref{fig:decision}-a), and  a tuple $t =$($N_1,2,N_2,3$). To decide whether to impute or delay missing value $t.a$ ($N_1$),  QUIP estimates the total imputation cost  in case it chooses to impute or to delay imputing $t.a$. The set of possible executions that may result for either
of the decisions are illustrated in the decision tree shown in 
 Figure~\ref{fig:decision}-b). Each  path of the
 tree corresponds to a possible outcome 
 based  on the decision to impute/delay imputing $t.a$. 
For instance, in path $p_5$, $t.a$ is imputed but fails the predicate in $o_1$, while in path $p_3$, $t.a$ is delayed and $t$ passes the predicates associated with $o_2$ and $o_3$, and reaches the imputation operator $\rho$, where  $t.a$ is imputed and evaluated in $\rho$ using predicate associate with $o_1$. 
The estimated imputation cost in the case of imputing (delaying) $t.a$, is the summation of the \textit{expected} imputation cost of all paths in the left (right) side of tree, i.e., $p_1,p_2,p_5,p_6$.  ($p_3,p_4,p_7,p_8$). 
The expected costs of various paths (shown in Figure~\ref{fig:decision}-c)  are computed 
as a weighted sum of imputations along the path,
where the weight corresponds to the probability of execution of that imputation. For instance, for path $p_1$, we impute $t.a$ with the probability of 1, and, subsequently impute
$t.c$ with the probability of $S_{o_1}S_{o_2}$.
Thus, the cost of path $p_1$ is 
$impute(a) + S_{o_1}S_{o_2}impute(c)$.

\noindent\textbf{Cost Model for Query Processing.}  Since join costs dominate query execution, QUIP estimates query processing
costs by the corresponding join costs.
Consider the same decision tree in Figure~\ref{fig:decision}-b). The expected query processing cost if we impute (delay)  $t.a$ 
is the sum of the expected query processing costs for  all the paths in the left (right) side of the tree.  
Figure~\ref{fig:decision}-d) lists the probability of each path, and also, its query processing cost. The probability is estimated based on selectivity of the predicates along the path, and the cost is estimated by
 summing  execution  cost  of execution of  operators along the path  incurred
 in processing  tuple(s) that are generated as a result of processing $t$. 
Take $p_6$ as an example. 
Its corresponding probability is  $S_{o_1}(1-S_{o_2})$ since $t$ passes $o_1$ but fails $o_2$. The estimated cost 
for processing $t$ (shown 
in Figure~\ref{fig:decision}-a) in operator $o_1$, denoted by $QP(o_1)$, is  $\mathcal{T}_{o_1}*TTJoin_{o_1}$ which is $0$ in this example since $o_1$ is a selection operator for which
 $\mathcal{T}_o$ is 1 and $TTJoin_{o_1} = 0$. 
The cost 
$QP(o_2) = \mathcal{T}_{o_1}\mathcal{T}_{o_2}*TTJoin_{o_2}$ since $o_2$ is a join operator and  $\mathcal{T}_{o_1}\mathcal{T}_{o_2}$ is the estimated number of join tests to perform in $o_2$.  

\vspace{-0.5em}
\section{Evaluation}
\label{sec:evaluation}

In this section we evaluate QUIP over two real data sets and one synthetic data set.  
Similar to ImputeDB, we implemented QUIP on top of SimpleDB~\cite{simpledb}, a teaching database used at MIT, University of Washington, and Northwestern University, among others. We did so, so that we can directly measure further improvements due to QUIP on ImputeDB query plans. We run our evaluations in a single-node machine with 16GB memory, Apple M1 Pro chip and 1TB flash storage.

\vspace{-0.4em}
\subsection{Data sets}

\noindent\textbf{UCI WiFi}. The first data set, \textsf{UCI-WiFi}, is collected from real WiFi connectivity data in UCI campus by Tippers~\cite{mehrotra2016tippers}, a sensor data management system. \textsf{UCI-WiFi} has three tables, \textsf{users}, \textsf{wifi} and \textsf{occupancy} with $4018$, $240,065$ and $194,172$ number of tuples, and totally $383,676$ missing values respectively.  \textsf{Wifi} records the continuous connectivity data of devices - that is, which  device is at which location in which time interval. \textsf{occupancy} has the occupancy (i.e., the number of people) of locations as a function of time.

\noindent\textbf{CDC NHANES}. We use the subset of 2013–2014 National Health and Nutrition Examination Survey (NHANES) data collected by U.S. Centers for Disease Control and Prevention (CDC)~\cite{cdcdata}. \footnote{We thank   ImputeDB~\cite{cambronero2017query} for providing this data set whose link can be found in ~\cite{QUIP, ImputeDB}. } \textsf{CDC} data set has three tables, \textsf{demo}, \textsf{exams} and \textsf{labs}, which are extracted from a larger complete CDC data set. \textsf{demo}, \textsf{exams} and \textsf{labs} have 10175, 9813, 9813 tuples, respectively, and all of them have 10 attributes. Among them, there are totally 24 attributes that contain missing values, whose missing rate ranges from $0.04\%$ to $97.67\%$, with total $81,714$ missing values.

\noindent\textbf{Smart Campus}. We used the SmartBench~\cite{gupta2020smartbench} simulator to generate synthetic sensor and semantic data based on seed data collected from a real system at the UCI campus using Tippers~\cite{mehrotra2016tippers}. In \textsf{smart-campus} data set, we generate 2 semantic tables, \textsf{location}, \textsf{occupancy}, 4 synthetic sensor tables, \textsf{WiFi}, \textsf{Bluetooth}, \textsf{Temperature}, \textsf{Camera}, as well as a \textsf{space} table and \textsf{user} table. \textsf{smart-campus} data set has totally approximately 2 million (1,892,500) tuples and 1.6 million missing values (1,634,720).  
The more detailed metadata for the above three data sets are shown in Section~\ref{subsec:evaluation}.

\vspace{-0.2em}
\subsection{Query Set}
We create three query workloads to evaluate QUIP, \textit{random} (with random selectivity), \textit{low-selectivity} and \textit{high-selectivity}. In each query workload, the majority of queries are SPJ-aggregate queries that contains \textit{select}, \textit{project}, \textit{join}, \textit{aggregate} (\textit{group by}) operations. SP queries are also included. Each query workload contains 20 queries. 


\vspace{-0.2em}
\subsection{Imputation Methods}
We chose four imputation approaches, two blocking, i.e., Top-k nearest neighbor~\cite{knn} (Knn) and XGBoost~\cite{chen2016xgboost,histogramboost}, and two non-blocking, i.e., histogram-based mean value imputation~\cite{cambronero2017query} and LOCATER~\cite{lin2020locater}. Among them, Knn, XGBoost and mean value imputations are widely used and their implementations are available in standard Python packages, such as \textsf{sklearn} or \textsf{xgboost}. LOCATER, which is an expensive time-series imputation approach,  imputes missing location and occupancy in \textsf{UCI-WiFi} and \textsf{Smart-Campus} data set. This technology has already been deployed and running in UCI in several real location-based applications for over one year, such as occupancy~\cite{lin2021t} and crowd  tracing~\cite{ucioccupancy} . Thus, we believe LOCATER is good fit for real testing in our live problem settings.

\begin{figure}[tb]
    \centering
    \vspace{-1em}
    \includegraphics[width=0.87\linewidth]{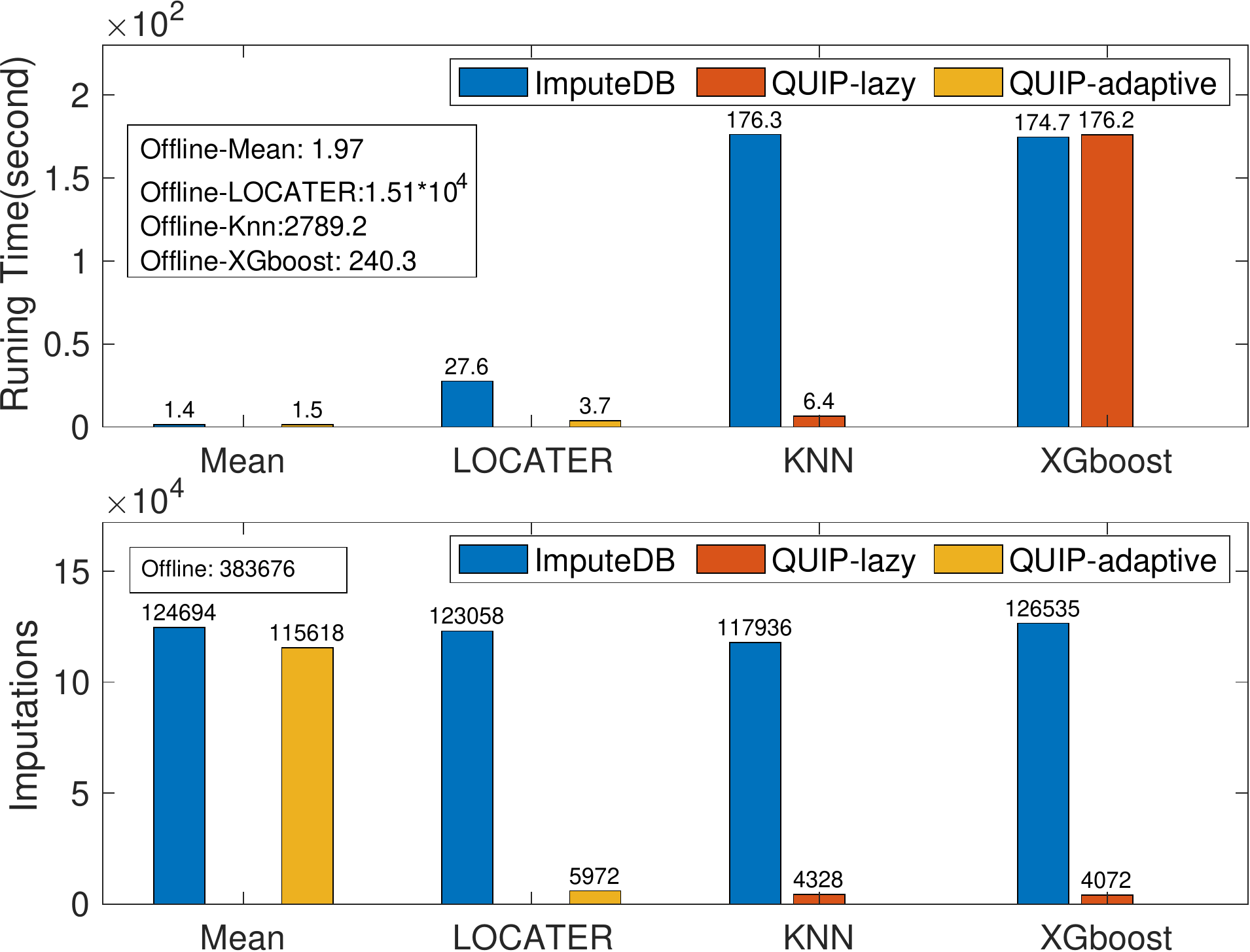}
    \vspace{-1em}
    \caption{\small UCI-WiFi Data Set.}
    \label{fig:wifi}
\end{figure}

\vspace{-0.3em}
\subsection{Strategies Compared}
We evaluated QUIP with two versions, \textit{QUIP-lazy} and \textit{QUIP-adaptive} as defined in Section~\ref{sec:decide}, and compared QUIP with \textit{Offline} and \textit{ImputeDB}. \textit{Offline} approach first imputes all missing values in the data set and then execute query processing. ImputeDB~\cite{cambronero2017query} uses  a parameter $\alpha$ between $0$ and $1$ to trade-off efficiency with 
quality. In particular,  ImputeDB drops tuples with missing
values in order to improve efffiency, though dropping tuples can result in reduced quality. It uses $\alpha$ as a parameter to explore such a tradeoff.  higher the value of $\alpha$, more tuples with missing value will be dropped, leading to reduced quality. Since we are not exploring tradeoff between quality and efficiency in this paper (our interest is in reducing the number of imputations without loss of quality), when comparing against ImputeDB we set the value of $\alpha$ to be 0 so as to prevent tuples with missing values from being dropped by ImputeDB. We note that ImputeDB's strategy for trading quality with efficiency can also be incorporated in QUIP though we do not explore such
a strategy in this  paper. We also compare QUIP with QuERy~\cite{altwaijry2015query}, and include the results in Section~\ref{subsec:evaluation}. 

\vspace{-1em}
\subsection{Results}
\begin{figure}[tb]
    \centering
    \includegraphics[width=0.88\linewidth]{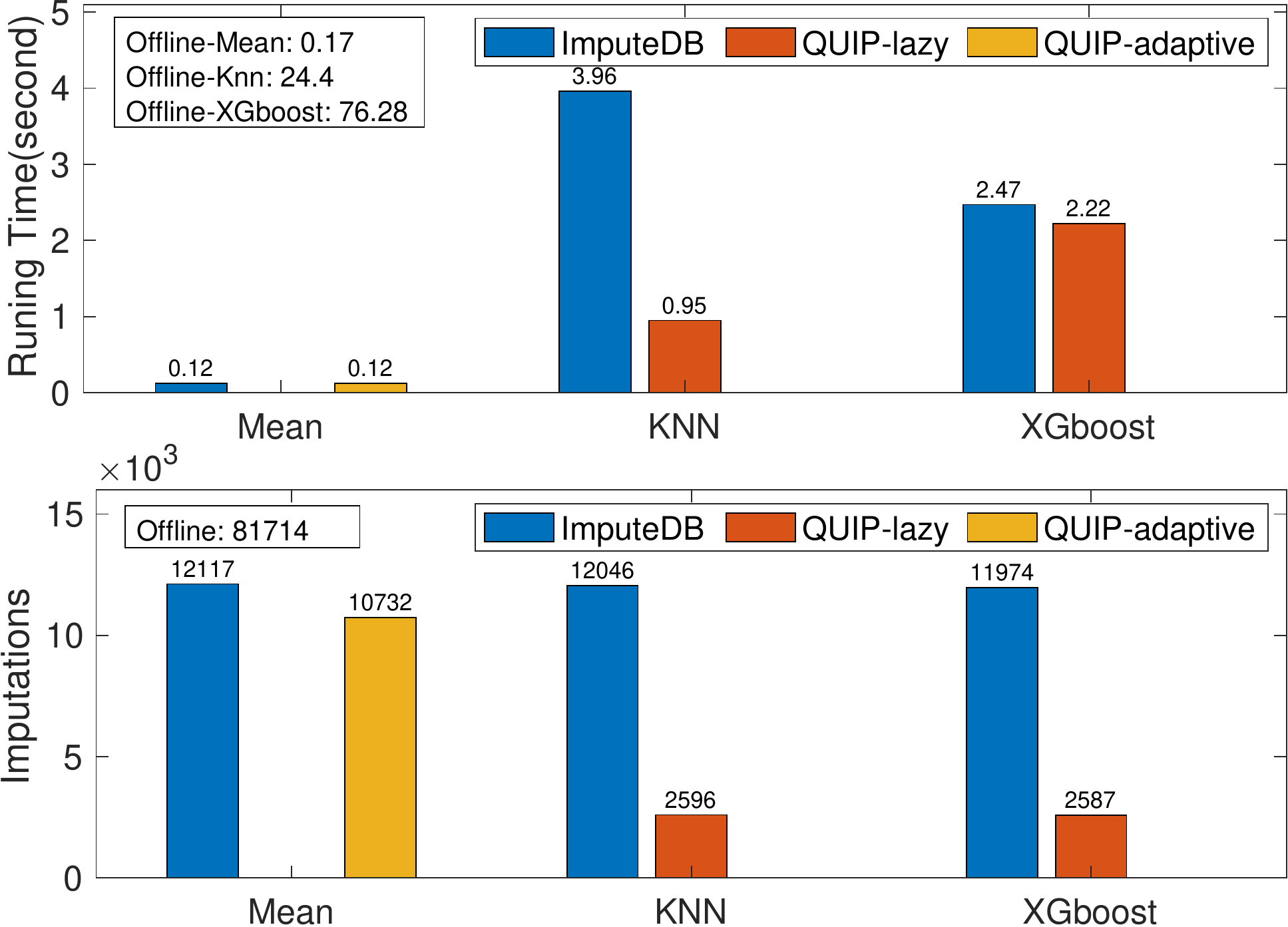}
    \vspace{-1em}
    \caption{\small CDC Data Set.}
    \label{fig:cdc}
\end{figure}

\vspace{-1mm}
\noindent\textbf{Experiment 1: Runtime \& Imputations.} 
In Figure~\ref{fig:wifi} and Figure~\ref{fig:cdc} we evaluate the runtime and number of imputations, i.e., the number of missing values that are imputed for \textit{Offline}, \textit{ImputeDB}, \textit{QUIP-lazy} and \textit{QUIP-adaptive} approaches using the \textit{random} query set. Among them, QUIP-lazy is applied on blocking imputations, i.e., KNN and XGboost while QUIP-adaptive is applied on non-blocking imputations, i.e., Mean and LOCATER. In \textsf{CDC} data set we only apply Mean imputation since LOCATER is not applicable. 
We make several observations. First,  both lazy and adaptive versions of QUIP achieves big savings in the number of imputation. It requires only   4.7\% and
and 21\% of imputations compared to  ImputeDB over UCI \textsf{UCI-WiFi}  dataset (in Figure~\ref{fig:wifi}) and  \textsf{CDC} dataset ( in Figure~\ref{fig:cdc}) respectively when expensive imputations are used, i.e., KNN, LOCATER and XGboost. Comparing with Offline approach, QUIP only imputes less than 1\% of missing values to answer the query. 
In contrast, when cheap imputations are used such as Mean imputation, QUIP tends to impute data  first since doing so will potentially save the query processing time by reducing temporary tuples, and thus has similar imputations as ImputeDB. 

Second, QUIP has similar performance with ImputeDB when cheap imputations (Mean) or  learning approaches with training and inference phases are used whose training time dominates the imputation costs (e.g., as in XGboost) are used. For such learning approaches, reducing imputation numbers would not make a big improvement if inference cost is negligible comparing with training time. However, it would be expected to make a difference if the learning approach whose inference cost is comparable to training, as in the case of  KNN and SVM (see Figure~\ref{fig:ML}).

Third, QUIP outperforms ImputeDB and Offline by around 20x and 1000x in \textsf{UCI-WiFi}, and the improvements are around 4x and 85x in \textsf{CDC} when costly non-blocking imputations (LOCATER) or blocking imputations with expensive inference (KNN) are used. 

The above observations demonstrate that QUIP by wisely delaying imputations can potentially achieve significant savings  leading to great improvement of query execution time when imputations with expensive inference time are used. 

\begin{figure*}[tb]
	\centering
	\begin{minipage}[t]{0.49\linewidth}
		\subfigure[\small CDC Data Set.]
		{\label{fig:cdcselectivity}\includegraphics[width=4.2cm,height=3.3cm]{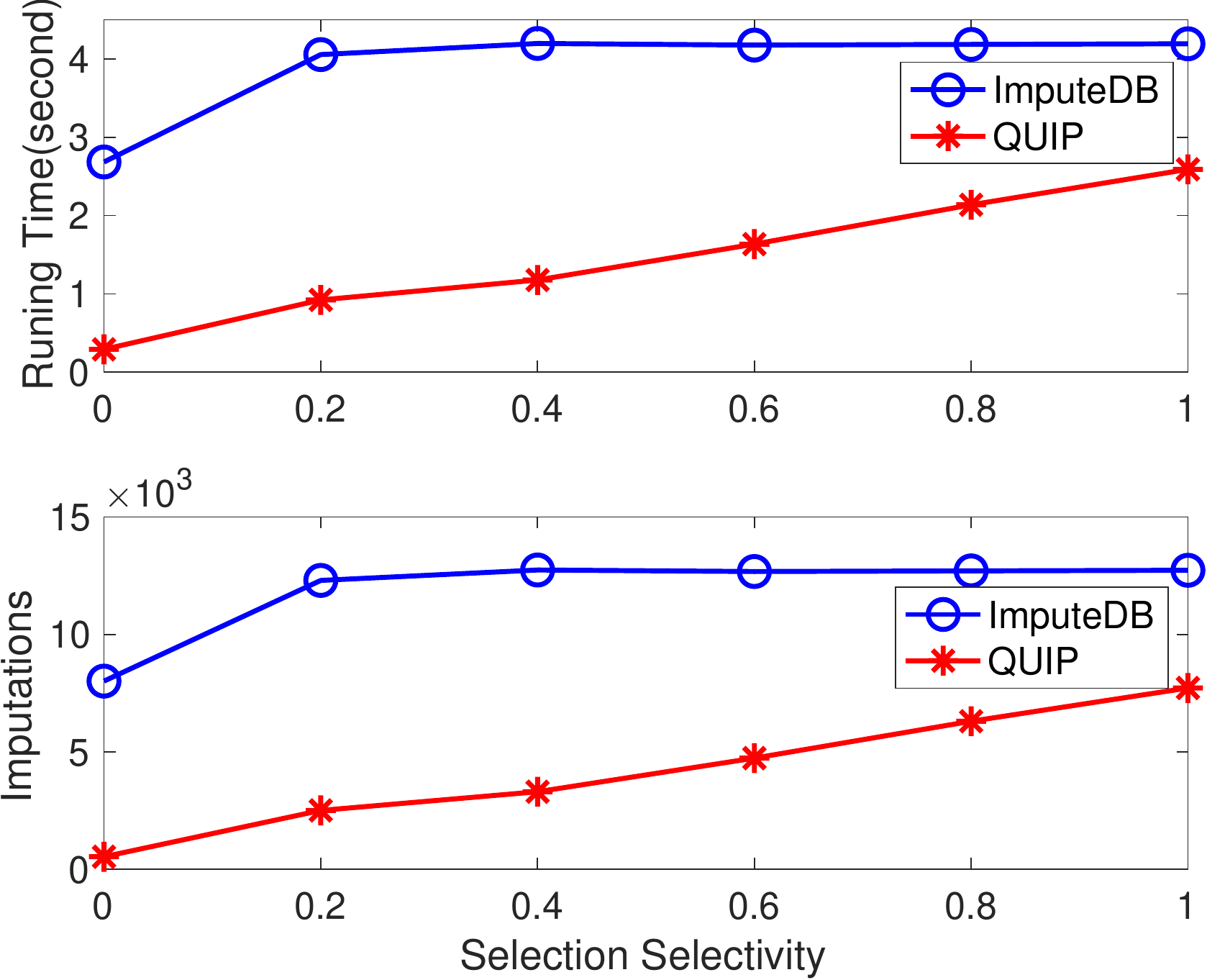}}
		\subfigure[\small WiFi Data Set.]
		{\label{fig:wifiselectivity}\includegraphics[width=4.2cm,height=3.3cm]{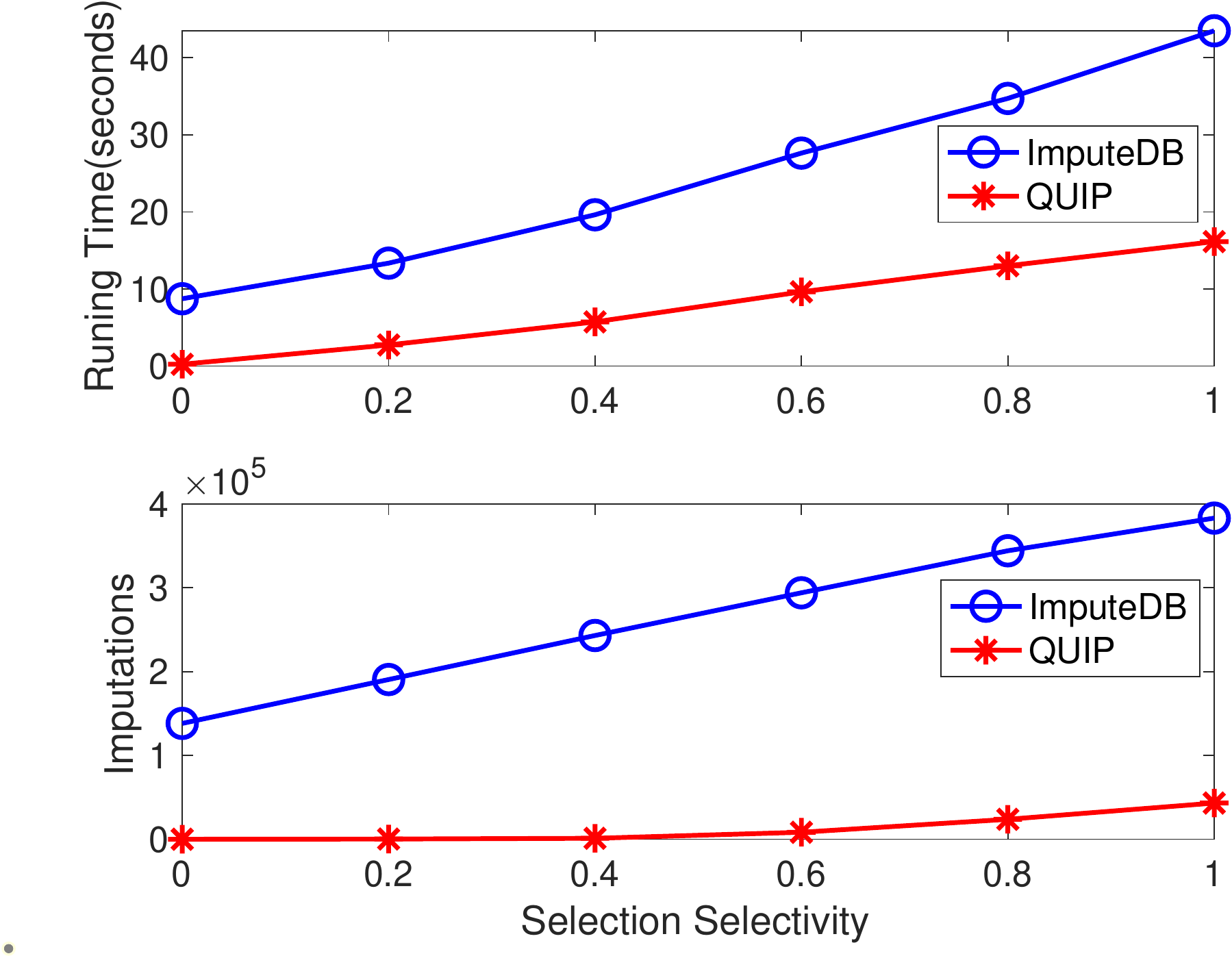}}
		\vspace{-1.3em}
		\caption{\small Selectivity Effects on Real Data Set.}
		\label{fig:realselectivity}
	\end{minipage}
	\begin{minipage}[t]{0.49\linewidth}
		\subfigure[\small Low Join Selectivity.]
		{\label{fig:acslowselectivity}\includegraphics[width=4.2cm,height=3.3cm]{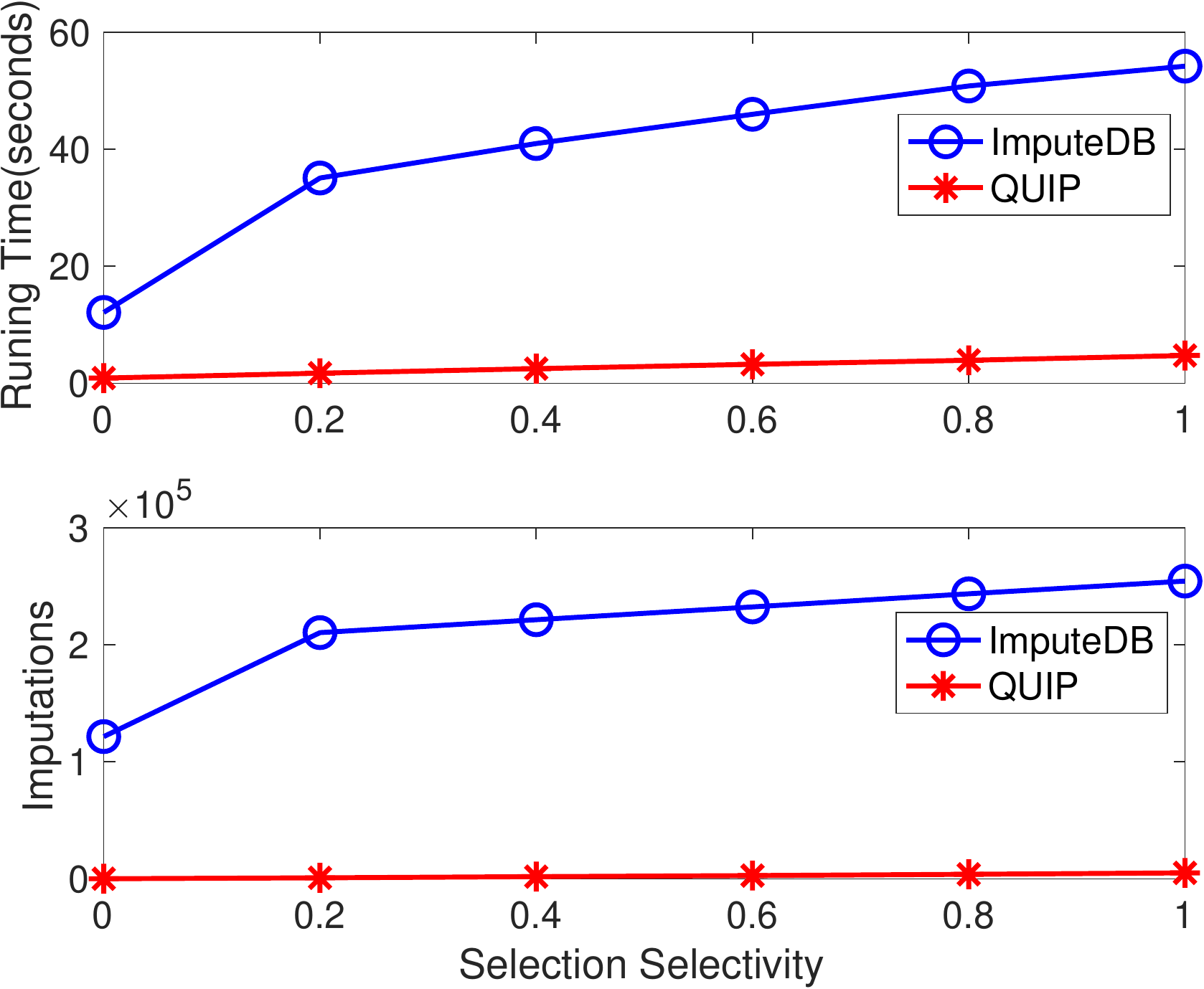}}
		\subfigure[\small High Join Selectivity.]
		{\label{fig:acshighselectivity}\includegraphics[width=4.2cm,height=3.3cm]{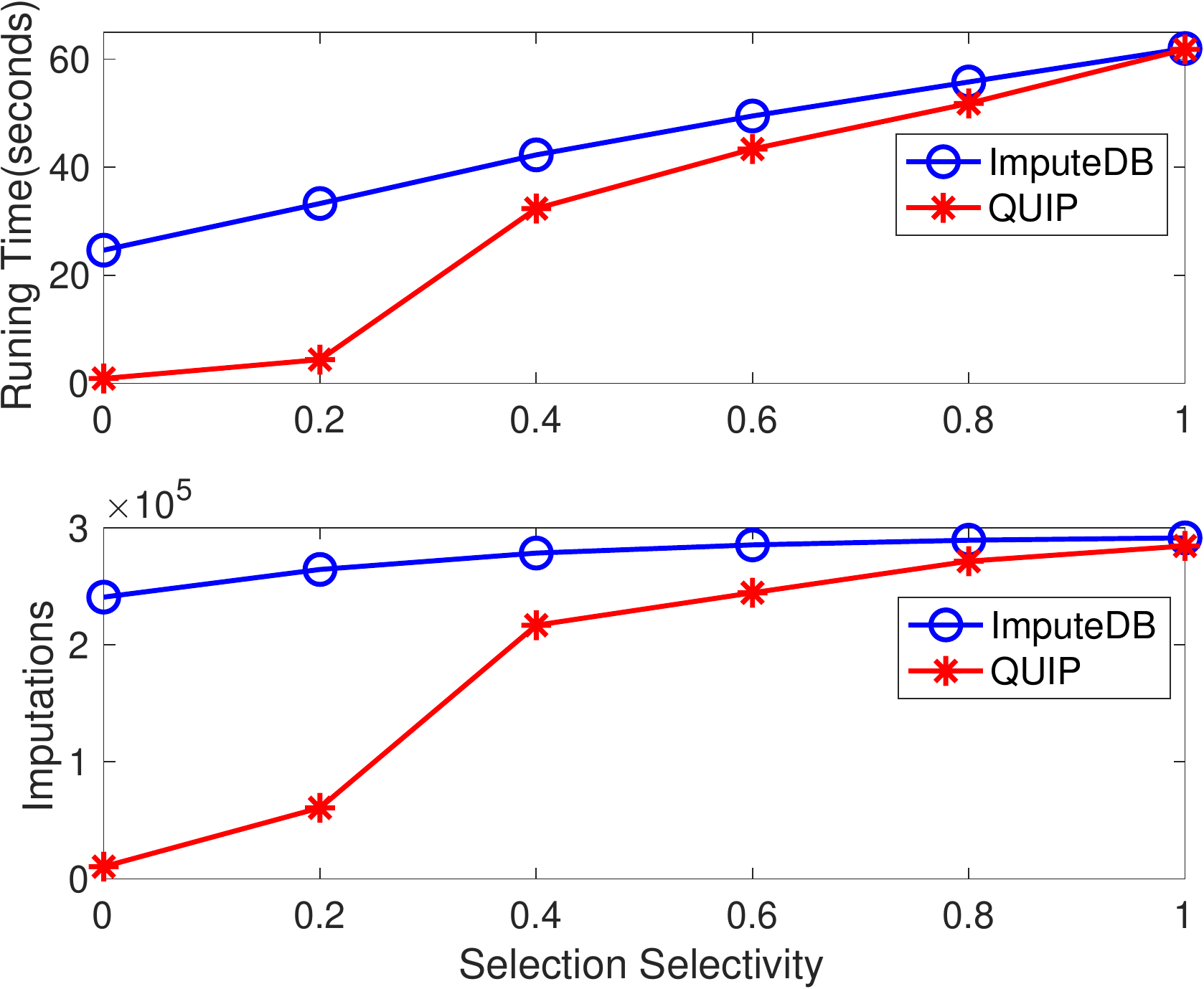}}
	\vspace{-1.3em}
	\caption{\small Selectivity Effects on Synthetic Data Set.}
	\label{fig:syntheticselectivity}
	\end{minipage}
	\vspace{-1em}
\end{figure*}

\vspace{0.5mm}
\noindent\textbf{Experiment 2: Quality of Query Answer.}
QUIP achieves  the same answers as the  eager strategy (that imputes
all missing data prior to executing the query) irrespective of whether the imputation strategy is  either blocking and non-blocking. ImputeDB, in contrast, may result in different answers when using a learning-based method. In ImputeDB, depending
upon where the imputation operator is placed, the model learnt by 
the imputation operator may differ based on the subset of data 
that is input into the imputation operator. As a result, 
the quality of the answer may vary. To compare 
 QUIP to ImputeDB in terms of quality 
under KNN imputation (which is blocking) using Symmetric-Mean-Absolute-Percentage-Error (SMAPE)~\cite{makridakis2000m3}. This is the
 quality metric used in ImputeDB \cite{cambronero2017query}. 
 We compute SMAPE as tuple-wise absolute percentage deviations for each query. 
On the \textsf{CDC} and \textsf{UCI-WiFi} data set, 
SMAPE value for for ImputeDB is 0\% to 4\% in \textsf{CDC} data set and  0\% to 3\% in \textsf{UCI-WiFi} data set.  In contrast, 
by design (and also validated experimentally) QUIP achieves 
a SMAPE of 0 in both data sets.

\vspace{0.5mm}
\noindent\textbf{Experiment 3: Query Selectivity Effects.} 
We first use the following query template to generate \textit{low-selectivity} and  \textit{high-selectivity} query workloads: 
\textsf{SELECT a, AVG(b) FROM $R_1, ..., R_n$ WHERE  [$Pred_J$] [$Pred_S$] GROUP BY a.}, where $Pred_J$ refers to the set of join predicates, $\{R_i.x = R_j.x\}$ among all relations, and $Pred_S$ refers to a set of selection predicates,  $\{R_{i}.a_i >= x_i\}$ by choosing a random attribute $a_i$ in every relation. 
We varies the selectivity of each selection predicate as 0, 0.2, 0.4, 0.6, 0.8, 1 by changing the operands $x_i$, and the selectivity of join predicate is set to be low and high by modifying the matching numbers of joined attribute values. KNN is applied in \textsf{CDC} data set, while in \textsf{UCI-WiFi} and  \textsf{Smart-Campus} data set, LOCATER is used to impute location and occupancy, and Mean-value is used to impute other missing values. 
In \textsf{CDC} and \textsf{UCI-WiFi} data set, we report the effect from selectivity of selection predicates in Figure~\ref{fig:realselectivity}, and the effects from both join and selection selectivity are evaluated in synthetic data set in Figure~\ref{fig:syntheticselectivity}. 

With  increasing  selectivity of selection predicates, both ImputeDB and QUIP have an increase in imputation numbers and running time, and QUIP has a considerably lower imputations and running overhead than ImputeDB at all selectivity levels. In \textsf{CDC} data set where join operations are relatively selective, QUIP tends to delay imputations in selection operators because join predicates will help eliminate lots of temporary tuples and thus reduce the number of imputations in the case where imputation overhead is costly than query processing. In \textsf{WiFi} data set whose join attributes have missing values, instead of imputing all join attributes values before join as ImputeDB does, 
QUIP delays  imputations due to following two reasons. First, imputations in join attributes can be saved if such delayed tuples can be dropped using filter set in VF list. Second, in a temporary tuple $t$ containing missing values in join attributes, among several join and selection predicates, instead of imputing all missing values in tuple $t$, it is possible that only a few imputations will eliminate $t$ because the imputed values failed their predicates stored in verify set in VF list. 

In synthetic data set, it is interesting to note that when join selectivity is high, QUIP tends to first impute missing values in selection operators because join predicates would not help eliminate imputations much in this case, and thus QUIP's performance is closer to ImputeDB when the selection predicates is less selective. When the join operators are highly selective, i.e., selectivity is low,  QUIP tends to delay imputing missing values in selection operators because low join selectivity would help eliminate most temporary tuples and thus save most imputations. In Figure~\ref{fig:acslowselectivity}, QUIP outperforms ImputeDB considerably in this case. 

\begin{table}[]
\small
    \centering
    \begin{tabular}{|c|c|c|c|c|}
    \hline
    & CDC & UCI-WiFi & SM-low & SM-high\\ \hline
    $\triangle Runtime$ & 0 & 67ms & 31ms & 73ms \\ \hline
    $\triangle |Temporary Tuples|$ & 0 & 124,711 & 58,311 & 137,826 \\ \hline
    $\triangle Imputations$ & 0 & 0 & 0 & 0 \\ \hline
    \end{tabular}
    \caption{Effect of Bloom Filter.}
    \vspace{-1.5em}
    \label{tab:bloomfilter}
\end{table}

\vspace{0.5mm}
\noindent\textbf{Experiment 4: Bloom Filter Effect.}
We evaluate the effect of bloom filter used in QUIP and report the results in Table~\ref{tab:bloomfilter} where SM stands for \textsf{Smart-Campus}. Specifically, let $\triangle Runtime$ be the difference of running time between QUIP and QUIP without bloom filter. Similarly, $\triangle|TemporaryTuples|$ and $\triangle Imputations$ measures the difference of number of temporary tuples and imputations. We observe that using bloom filter would not help reduce imputations but help save the number of temporary tuples and thus reduce the query processing time. Note that it only works when the join attributes have missing values, which exist in \textsf{UCI-WiFi} and \textsf{Smart-Campus} data set. If there are no missing values under join attributes, such as \textsf{CDC} data set,  QUIP will not execute outer join and thus bloom filter will not be applied.

\begin{figure}
\subfigure{\includegraphics[width=4.1cm,height=3.3cm]{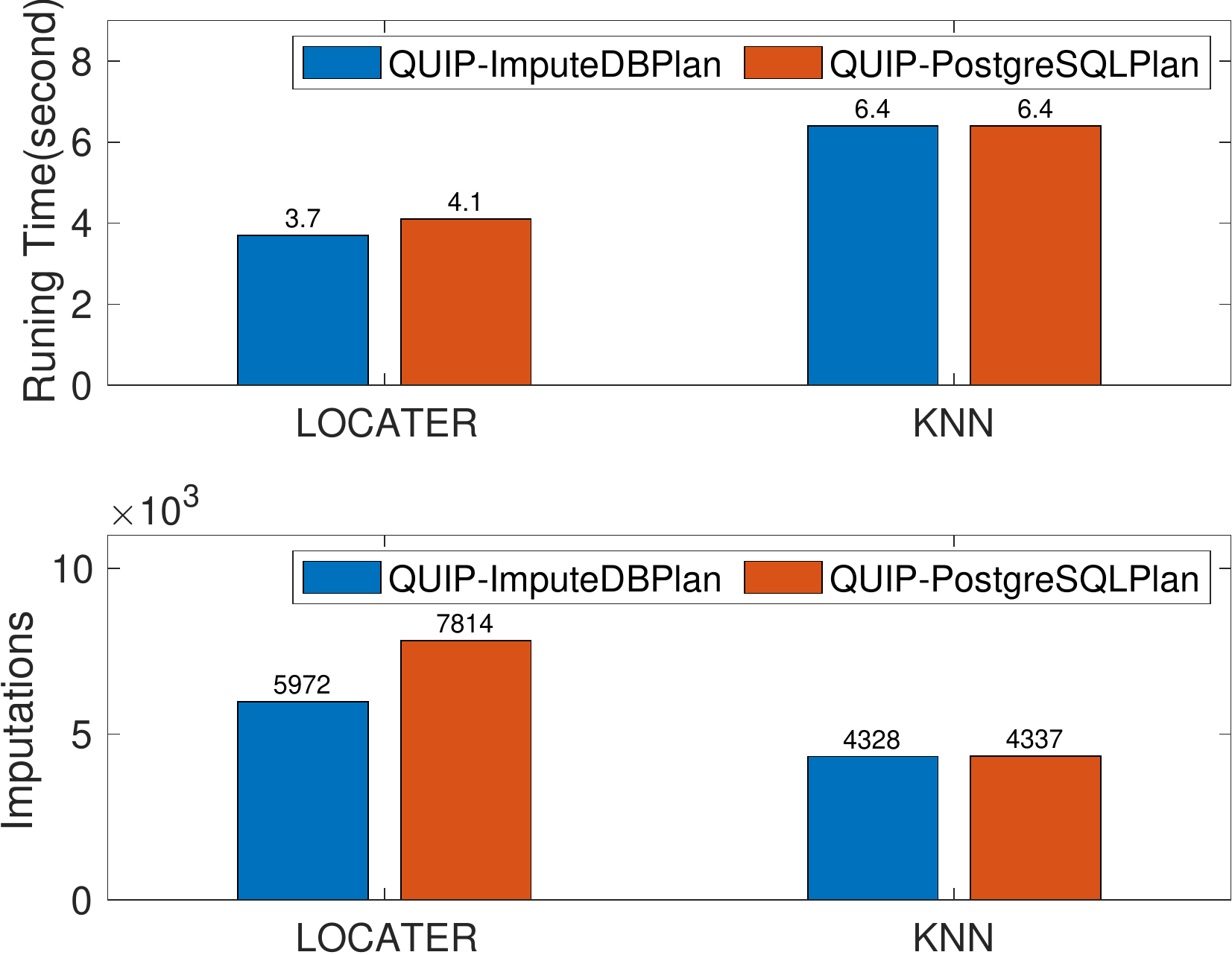}}
	\subfigure{\includegraphics[width=4.1cm,height=3.3cm]{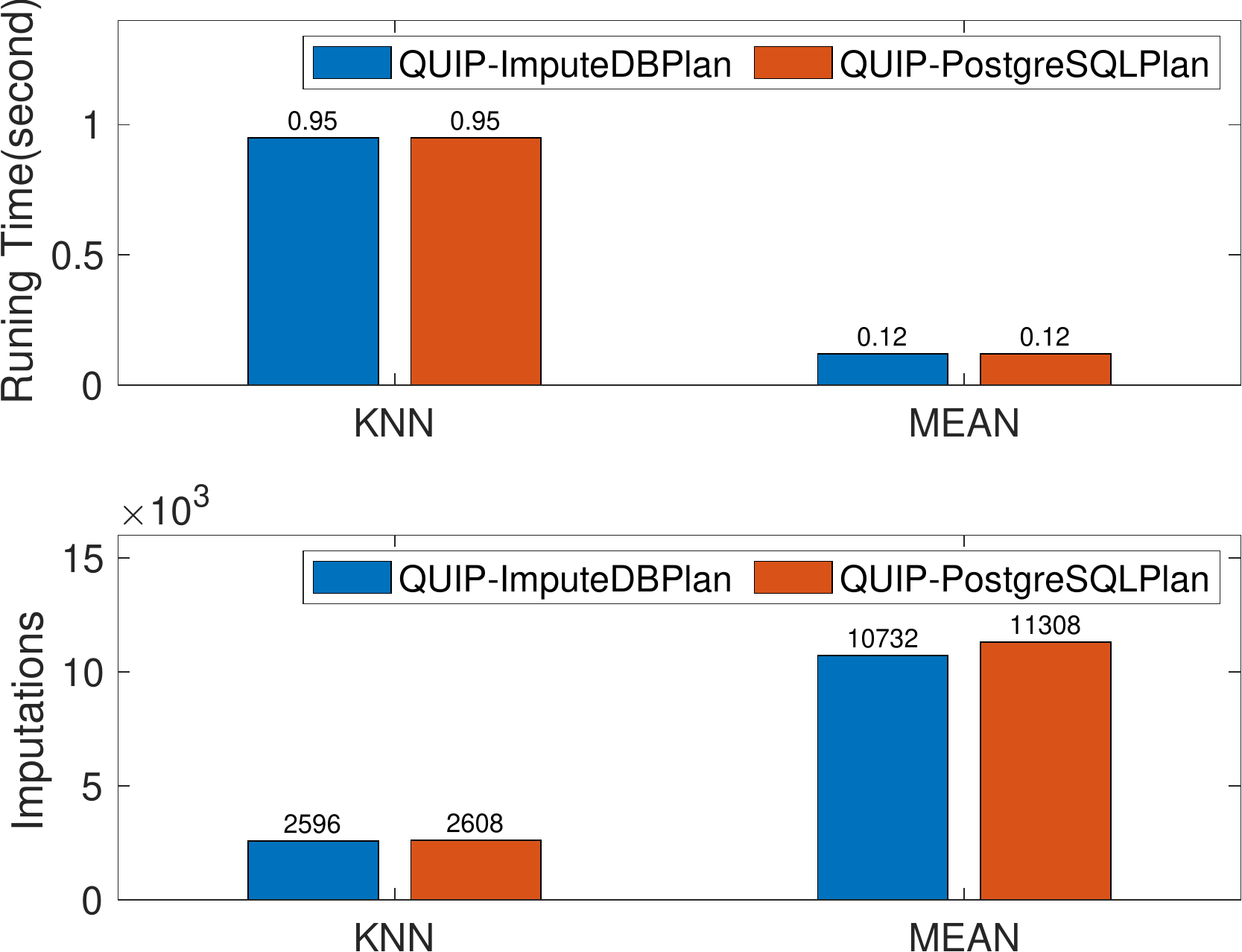}}
	\vspace{-1.5em}
\caption{\small QUIP with Different Query Plans.} 
\label{fig:plan}
\end{figure}

\vspace{0.5mm}
\noindent\textbf{Experiment 5: QUIP with Different Query Plans.} 
We compare the performance of QUIP that takes different query plans as input in real data sets in Figure~\ref{fig:plan}. In particular, QUIP takes the query plan from ImputeDB and PostgreSQL on \textit{random} query set. We observe that the performance of the lazy strategy of QUIP is not affected by the query plan it takes when blocking imputations (e.g., KNN) are used. This is because whatever the query tree is, all the missing values will be delayed in selection or join operator and imputed at imputation operator $\rho$ that is on the top of query tree, and thus lead a stable performance. We also observe that when taking different plans for non-blocking operators, such as LOCATER, the plan generated by ImputeDB is better, which demonstrates the effectiveness of ImputeDB. But taking the plan by PostgreSQL also leads to an acceptable performance, which shows the robustness of QUIP to different underlying query plans.

\vspace{-1.2em}
\section{Conclusion}
\vspace{-1mm}
\label{sec:conclusion}

This paper studies query-driven missing value imputation and proposes QUIP, a technique to intermix query processing and missing value imputation to minimize query overhead by taking a reasonable good query plan as input. Specifically, QUIP co-optimizes imputation cost and query processing cost, and proposes a new implementation based on outer join to preserve missing values in query processing. 
Real experiments shows that QUIP outperforms the state-of-the-art technique ImputeDB 2 to 10 times and achieves order of magnitudes improvement over standard offline approach. 

 
\newpage

\section{Appendix}
\label{sec:appendix}

\subsection{Decision Functions for Blocking Imputations}
\label{subsec:blocking}

In the context of blocking-based imputations, we need to differentiate between learning-based strategies that have a separate learning and inference (imputation) stages versus other strategies that do not have a learning phase. 

Let us first consider blocking learning-based strategy with dominant learning part and its imputation cost dominates query processing costs. In this case, QUIP will take an \textit{eager} strategy, where QUIP will impute missing values of the attribute(s) associated with the operator right away prior to evaluation of the operator. Such a strategy is exactly the one that ImputeDB takes. 

For those learning-based imputations where learning cost is not dominant, or other blocking imputations which have high cost (e.g., rule-based approaches~\cite{song2015enriching,fan2012towards}), it is advantageous to use a \textit{lazy} strategy where QUIP always delays imputing until the tuple with 
the missing value reaches  the top of query tree, i.e., imputation is performed inside the imputation operator $\rho$ in Section~\ref{sec:algorithm}. 
Note that in case of learning-based imputations where learning cost is not dominant and inference cost is comparable, the learning phase can be done eagerly and inference part becomes non-blocking, and thus our designed decision functions could be applied on inference. 
Both strategies can be simulated in QUIP 
by setting the decision function to always impute (eager) or to always delay (lazy). 

\subsection{Cost Models in Decision Function}
\label{subsec:costmodel}
\begin{figure}[tb]
    \centering
    \vspace{-1em}
    \includegraphics[width=0.95\linewidth]{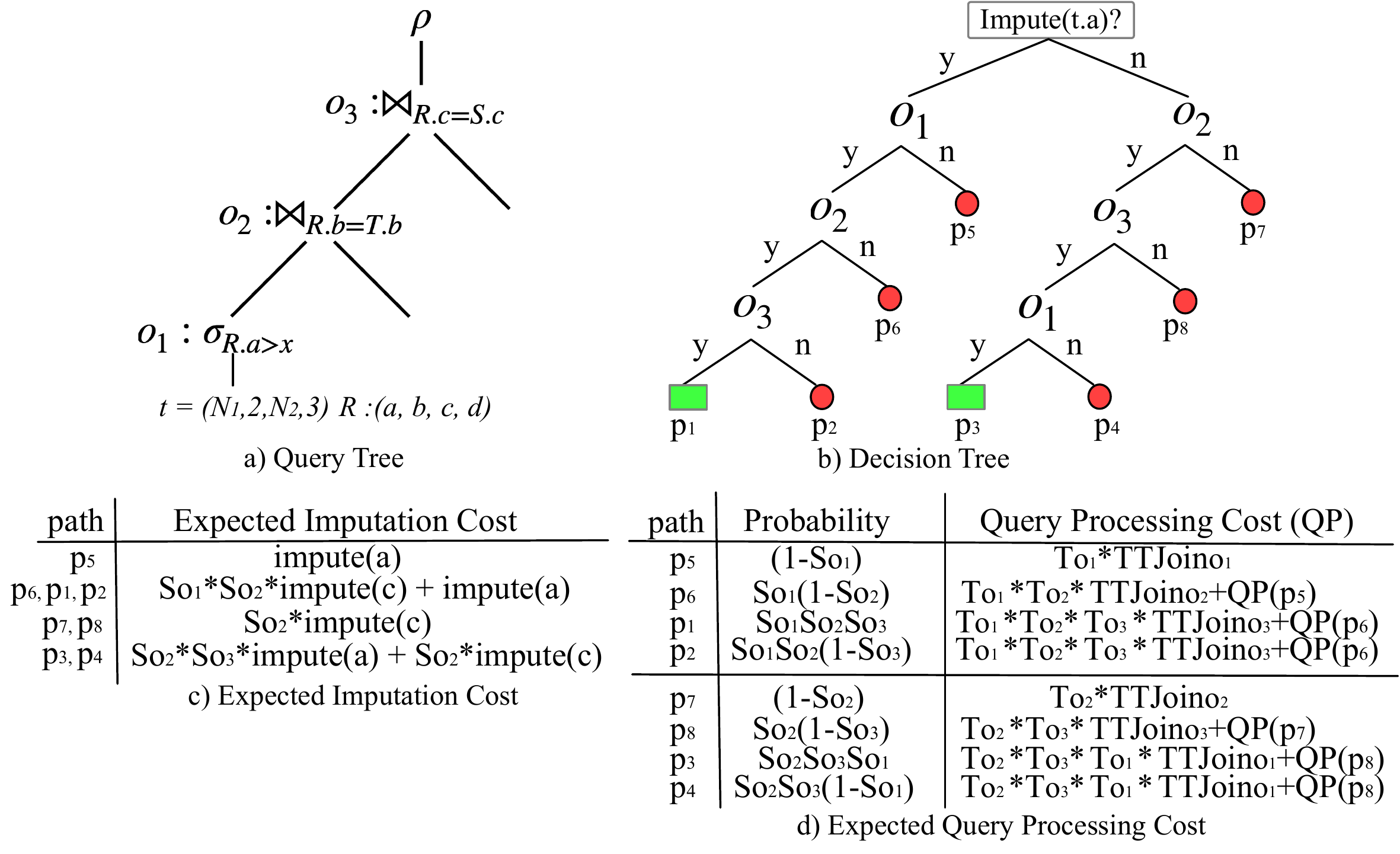}
    \vspace{-1em}
    \caption{\small Decision Function Example.}
    \label{fig:decision1}
\end{figure}

\noindent\textbf{Cost Model for Imputations.} 
For a tuple $t$ with missing value in attribute $a$ received by operator $o$ in query tree, 
$df(a,o)$ computes the expected imputation cost in the case of impute or delay by building a decision tree, as shown in Figure~\ref{fig:decision1}-b). Consider a tuple $t = (N_1,2,N_2,3)$ received in $o_1$, decision tree contains $o_1$ and its  downstream operators before imputation operator, i.e., $o_1,o_2,o_3$. 
Let each root-to-leaf path in decision tree be $p=\{decision,o_i,...,o_j,result\}$. For simplicity, we denote such a path by $p = \{o_i,...,o_j\}$. Each of such path $p$ corresponds to a possible evaluation result of $t$. Those paths with \textit{result} as red colors means tuple $t$ is eliminated while the green color means $t$ passes all predicates. Each edge from a operator $o$ denotes the \textit{evaluation result} of $t$, i.e., $y$ means $t$ passes $o$ while $n$ means $t$ fails $o$. For instance, $p_5$ corresponds to the case where $t.a$ is imputed and fails the predicate associate with $o_1$, while $p_3$ means that the imputation of $t.a$ is delayed (and thus cannot be evaluated in $o_1$), and $t$ passed the predicates in $o_2,o_3$, then $t.a$ is forced to be imputed in imputation operator and be re-evaluated using $\sigma_{R.a>x}$ associated with $o_1$. 

Let $E[IMP(impute)]$ and $E[IMP(delay)]$ be the expected imputation cost of tuple $t$ based on imputing and delaying $t.a$. \\$E[IMP(impute)]$ ($E[IMP(delay)]$) is computed by summarizing the expected cost of paths where $t.a$ is imputed (delayed) in decision tree. For instance, $E[IMP(impute)] = E[IMP(p_1)] + E[IMP(p_2)] + E[IMP(p_5)] + E[IMP(p_6)]$.  The expected cost of a path $p$, $E[IMP(p)] \\=  \sum_{o_i\in p \wedge t.a_i=missing}impute(a_i)*Prob^{I}(o_i)$, where $a_i$ is the attribute of $t$ to be evaluate in $o_i$ and $Prob^{I}(o_i)$ is the probability that $t$ will reach operator, $Prob^{I}(o_i) = \prod_{o_c\in f(o_i)\wedge t.a_c=missing}S_{o_i}$, where $f(o_i)$ is the ancestors of $o_i$ in decision tree. If $f(o_i) = \emptyset$, then $Prob^{I}(o_i) = 1$. Figure~\ref{fig:decision1}-c) lists the expected imputation cost of all paths. Taking $p_3$ for example, if $t.a$ is delayed and $t$ could reach $o_3$ with probability $S_{o_2}$, the expected imputation cost for $t.c$ ($N_2$) is  $S_{o_2}*impute(c)$. Similarly, if $t$ reach the imputation operator and $t.a$ be imputed there with probability $S_{o_2}*S_{o_3}$, then the expected imputation cost of $t.a$ is $S_{o_2}*S_{o_3}*impute(a)$.

\noindent\textbf{Query Processing Cost.} In SPJ query, we assume the join cost dominates the overall query processing cost, and thus we use the estimation of join cost to approximate query processing time. 
Based on the same decision tree built for deciding imputing $t.a$ or not, let $E[QP(impute)]$ and $E[QP(delay)]$ be the expected query processing cost based on imputing or delaying $t.a$. Similarly, $E[QP(impute)]$ is summation of the costs of the paths where imputing $t.a$ is decided, i.e., $E[QP(impute)] = E[QP(p_1)]+E[QP(p_2)]+E[QP(p_5)]+E[QP(p_6)]$. The expected query processing cost of a path $p$ is the summation of the expected cost in each operator in this path, i.e., $E[QP(p)] = \sum_{o_i\in p}(QP(o_i)*Prob^{E}(o_i))$, where $QP(o_i) = (\prod_{o_c\in f(o_i)\cup o_i}\mathcal{T}_{o_i})*TTJoin_{o_i}$, and $Prob^{E}(o_i) = (\prod_{o_c\in f(o_i)\cup o_i}(I(o_c)S_{o_c}\\+(1-I(o_c)(1-S_{o_c})))$. $I(o_c)$ is the indicator function that if $t$ passes the predicate associated with $o_c$, then $I(o_c)=1$, otherwise $I(o_c)=0$.  Figure~\ref{fig:decision1}-d) lists the expected imputation time cost for all paths in decision tree. Taking $p_6$ for example, its probability is $S_{o_1}(1-S_{o_2})$ since $t$ passes $o_1$ but fails $o_2$. The estimated query processing time in $o_1$, $QP(o_1)$, is simply $\mathcal{T}_{o_1}*TTJoin_{o_1}$ which is $0$ in this example since $o_1$ is a selection operator, while 
$QP(o_2) = \mathcal{T}_{o_1}\mathcal{T}_{o_2}*TTJoin_{o_2}$ since $o_2$ is a join operator and $\mathcal{T}_{o_1}\mathcal{T}_{o_2}$ is the estimated number of join tests to perform in $o_2$. (Note that as we mentioned before, $\mathcal{T_o}$ is 1 if $o$ is a selection operator.) 

\noindent\textbf{Decision Making.}
Based on the estimated imputation and query processing cost of  \textit{impute} and \textit{delay} decisions for missing value $t.a$, decision function will impute $t.a$ if $(E(IMP(impute) - E(IMP(delay))) + E(QP(impute)) - E[QP(delay)] < 0$; otherwise, decision function will delay imputing $t.a$. 

\begin{figure}[tb]
	\centering
	\includegraphics[width=0.95\linewidth]{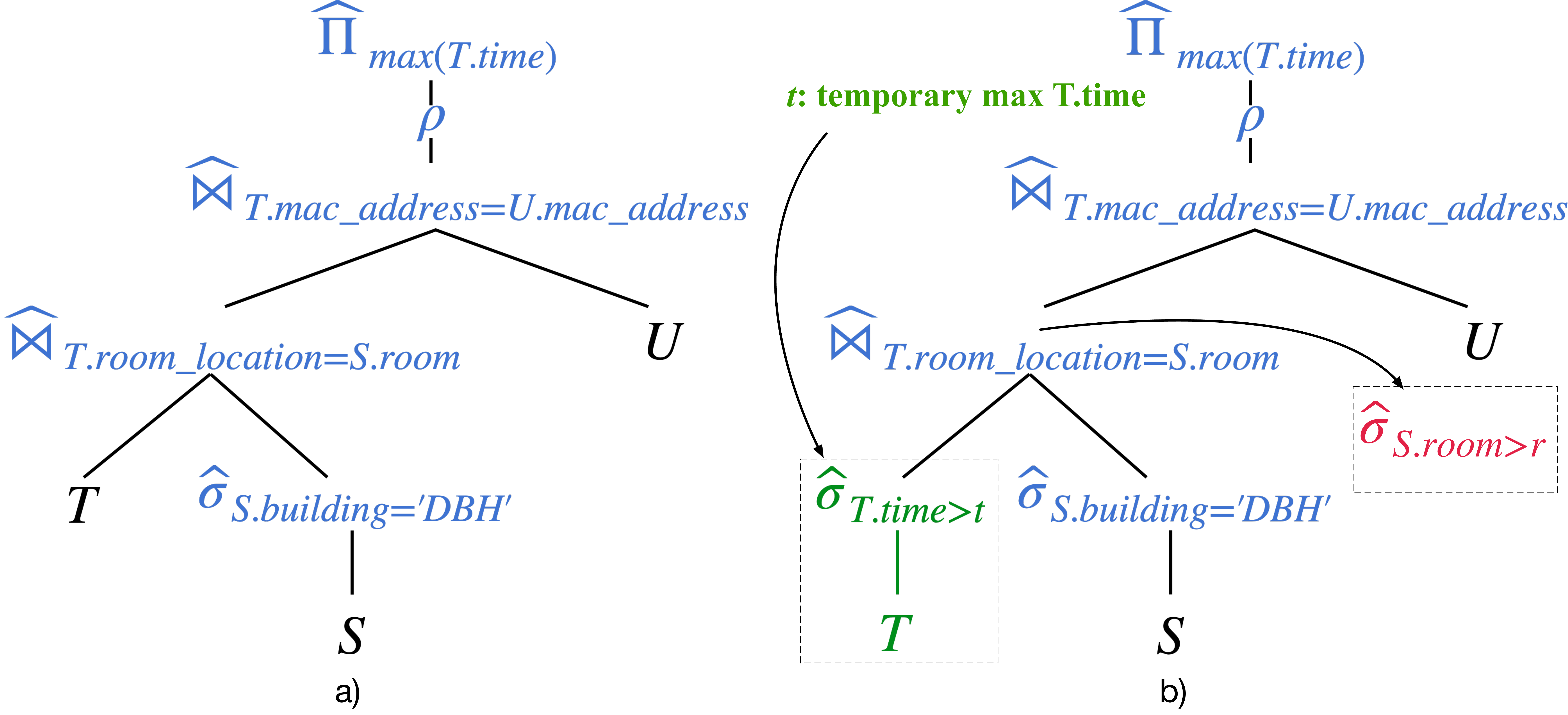}
	\caption{Optimization of max operator.}
	\label{fig:aggregate}
\end{figure}

\subsection{Extensions to Other Operators}
\label{subsec:extensions}
In this section, we extend QUIP to handle aggregate, union, set minus operators and nested query. In addition, QUIP uses an optimization technique to speed up \textit{max} and \textit{min} query, and we show in Section~\ref{subsubsec:minmax} that the proposed optimization can improve MIN/MAX query around 2x to 4x in different query workloads.

\begin{figure}
\raggedright
\texttt{
\hspace{-0.5em}\textcolor{blue}{max} T.time \\ \textcolor{blue}{FROM} Trajectories as T, Space as S, User as U \\
\textcolor{blue}{WHERE}
T.mac$\_$address = U.mac$\_$address
\textcolor{blue}{AND} \\
T.Room$\_$location = S.room
\textcolor{blue}{AND}
S.building = ‘DBH’ 
}
\vspace{-1em}
    \caption{Aggregation Query}
    \label{fig:query1}
\end{figure}

\noindent\textbf{Aggregate Operators.}
QUIP can be easily extended for aggregate operators, \textsf{max}, \textsf{min}, \textsf{count}, \textsf{sum}, \textsf{avg}, \textsf{group by} by adding them in the top of query plan tree. Among them, QUIP specially optimizes for \textsf{max} and \textsf{min} operators, which are two of commonly  used operators for data analysis. Consider aggregation query in Figure~\ref{fig:query1} which seeks to find the latest time stamp of WiFi connectivity events that happens in 'DBH' building. The query plan tree is shown in  Figure~\ref{fig:aggregate}-a correspondingly. During query processing, QUIP maintains a temporary maximum time $t$ under $T.time$ it has seen so far, and creates a selection operator $\widehat{\sigma}_{T.time>t}$. QUIP then pushes this operator to the leaf as shown in Figure~\ref{fig:aggregate}-b to filter out all the tuples in relation $T$ whose \textit{time} value is less than or equal to the current max time $t$. Likewise, for \textsf{min} operation, \textsf{select min(T.time)}, QUIP would create a selection operator $\widehat{\sigma}_{T.time<t}$ where $t$  corresponds to temporary minimum value for  $T.time$. 
Note that tuples filtered by this selection condition do not affect the query result since they cannot be the part of query answer. 
To see why this optimization can be powerful, assume that the first temporal maximum value we see happens to be a large number, which is close to the final maximum. In such a case, the selection operator we added   $\widehat{\sigma}_{T.time>t}$ will be very selective which could help filter most tuples in relation $T$. 

To complete this technique, we need to specify when to maintain this temporal value and where to put the introduced selection operator. 
First, the temporal value QUIP maintains should be valid, which means that it should pass all the selection and join predicates. QUIP maintains this value just above any selection and join operators close to the root. 
Second, we push down the  selection operator introduced similar as many query optimizers do to potentially remove more tuples. Thus generally, QUIP will push down this selection operator to the corresponding leaf node in the tree except in  one special case. If the aggregate attribute $a$ is in a relation $R$ which is \textit{blocked} in query processing, then the introduced selection operator related with $a$ will be placed just above the first join operator associated with $R$. As an instance, if we want to find the room with maximum room ID, i.e., $max S.room$, and a hash joined is implemented in join operator $\widehat{\bowtie}_{T.room\_location = S.room}$. Assume that the right relation $S$ is \textit{build} relation, which will be used to build a hash table. If we still put the introduced selection operator $\widehat{\sigma}_{S.room>r}$ at the leaf under this join operator, it will not help eliminate tuples because the first valid tuple in $S$ relation will be popped up only after all tuples in $S$ have been scanned to build a hash table. In this case, QUIP will place $\widehat{\sigma}_{S.room>r}$ above $\widehat{\bowtie}_{T.room\_location = S.room}$ as shown in Figure~\ref{fig:aggregate}. Note that as time goes by, the introduced selection operator will become more selective as the maintained temporal value should be closer to real one. We show in our experiments that this simple optimization for \textsf{max} and \textsf{min} query is powerful and will improve the query execution time up to 2x to 4x in our tested query sets. 

\begin{table*}[]
\small
    \centering
    \begin{tabular}{|c|c|c|c|c|c|}
    \hline
          \multicolumn{2}{|c|}{demo (10175 rows)} &\multicolumn{2}{c}{labs (9813 rows)} &\multicolumn{2}{|c|}{exams (9813 rows)} \\ \hline
         Attr & Missing &   Attr & Missing & Attr & Missing \\ \hline
         age\_months & 93.39\%  & albumin &      17.95\% & arm\_circumference &       5.22\% \\
        age\_yrs &  0.00\% & blood\_lead &      46.86\% & blood\_pressure\_secs &       3.11\% \\
        gender &    0.00\% & blood\_selenium &      46.86\% & blood\_pressure\_systolic &      26.91\% \\
        id &  0.00\% & cholesterol &      22.31\% &   body\_mass\_index &       7.72\% \\
        income &  1.31\%& creatine &      72.59\% &   cuff\_size &      23.14\% \\
        is\_citizen &       0.04\% & hematocrit &      12.93\% & head\_circumference &      97.67\% \\
        marital\_status &      43.30\% & id &       0.00\% & height &       7.60\% \\
        num\_people\_household &       0.00\% & triglyceride &      67.94\% & id &       0.00\% \\
        time\_in\_us &      81.25\% & vitamin\_b12 &      45.83\% & waist\_circumference &      11.74\% \\
        years\_edu\_children &      72.45\% & white\_blood\_cell\_ct &      12.93\% & weight &       0.92\% \\
        \hline
    \end{tabular}
    \caption{CDC NHANES Data Set.}
    \label{tab:cdc}
\end{table*}

\begin{table}[]
\small
    \centering
    \begin{tabular}{|c|c|c|c|c|c|}
    \hline
          \multicolumn{2}{|c|}{users (4018 rows)} &\multicolumn{2}{c}{wifi (240,065 rows)} &\multicolumn{2}{|c|}{occupancy (194,173 rows)} \\ \hline
         Attr & Missing &   Attr & Missing & Attr & Missing \\ \hline
         name &  0.00\%   & start$\_$time &  0.00\%  & lid &  0.00\%  \\
        mac$\_$addr &  19.95\%  & end$\_$time &  0.00\%  & start$\_$time &  0.00\%  \\ 
        email &  0.00\%  & lid &  51.38\%  & end$\_$time &  0.00\%  \\ 
        group &  89.77\%  & duration &  0.00\%  &  occupancy &  71.17\%  \\ 
          & & mac$\_$addr &  0.00\%  & type &   61.50\%  \\
        \hline
    \end{tabular}
    \caption{UCI-WiFi Data Set.}
    \label{tab:uci-wifi}
\end{table}

\begin{table}[bt]
\small
    \centering
    \begin{tabular}{|c|c|c|c|c|c|c|}
    \hline
     &  \multicolumn{2}{|c|}{\# of Imputations} & \multicolumn{2}{|c|}{Running Time(ms)} & |RT| \\ \hline
    Query & QUIP & QUIP- & QUIP & QUIP- & QUIP \\ \hline
    CDC-Q6 & 81 & 81 & 102 & 104 & 0 \\ 
    CDC-Q7 & 9 & 882 & 127 & 376 & 9806 \\ 
    CDC-Q8 & 241 & 1781 & 197 & 732 & 8823 \\ 
    UCI-WiFi-Q5 & 983 & 1672 & 3461 & 4971  & 53755 \\
    UCI-WiFi-Q6 & 747 & 3464 & 2873 & 7045 & 34988 \\ 
        \hline
    \end{tabular}
    \caption{MAX/MIN Optimizations.}
    \label{tab:maxmin}
\end{table}

\begin{figure}[tb]
	\centering
	\includegraphics[width=0.5\linewidth]{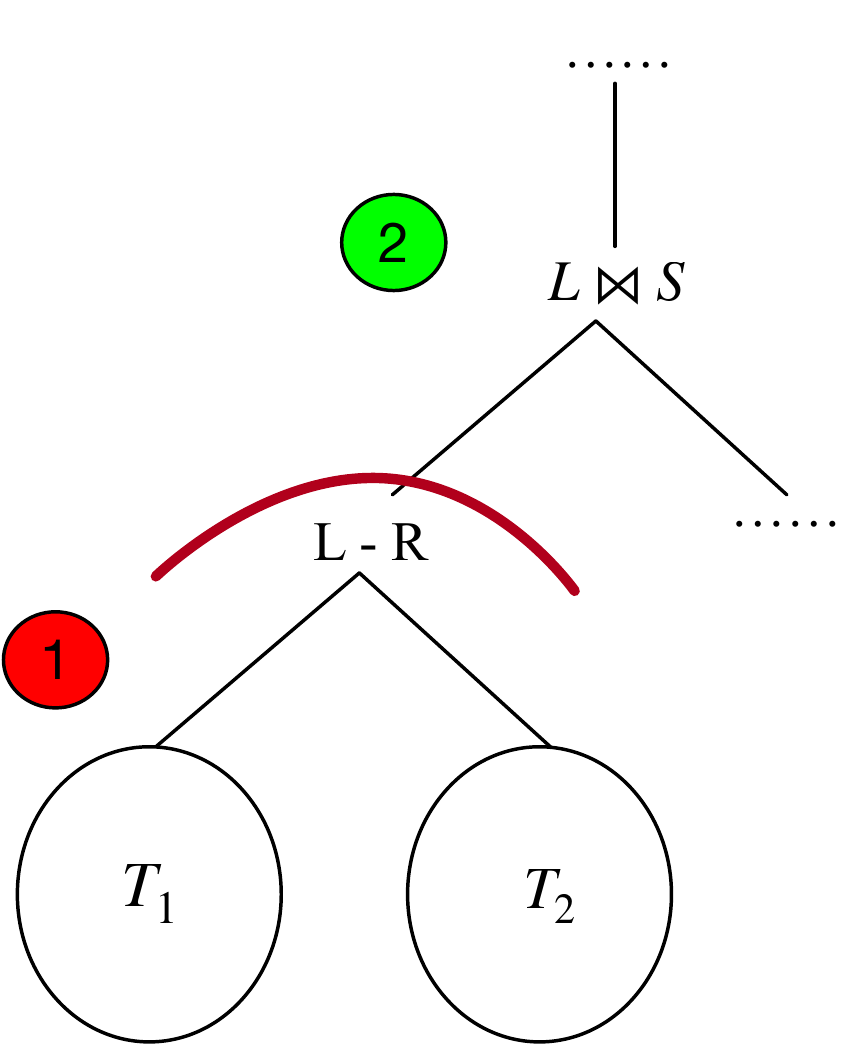}
	\caption{Set Minus Operator.}
	\label{fig:setminus}
\end{figure}

\noindent\textbf{Union Operator $\cup$.} Consider a union operation $L\cup R$. We modify union operator based on the logic in Figure~\ref{fig:operator}-b). For tuples received by union operator $L\cup R$ in both relations $L$ and $R$, we first apply \textit{filter} by using filter set in VF list in Figure~\ref{fig:vf}. Tuples that failed filter test will be dropped, and then for the other tuples, QUIP calls decision function to decide for all attributes, which are in $L$ and $R$ as well as appear in the given query predicates, whether to impute or delay imputations for missing values in the received tuples. For any missing value $v$ in the received tuples, if the decision function decides to impute $v$, QUIP will call \textit{verify} operation to evaluate the imputed value. If decision function decides to delay imputation, we just preserve the corresponding missing value in the tuple. A tuple whose all imputed values passed verification test will be forwarded above to next operator in query tree. $L\cup R$ will simply return  all tuples in $L$ and $R$ that passed filter and verify tests. 

\noindent\textbf{Set Minus Operator $-$.} Consider a set minus (also known as \textit{EXCEPT} in PostgreSQL) operation $L-R$ that returns tuples in relation $L$ but not in $R$. Set minus operator is a blocking operator for QUIP, and all the tuples returned by set minus operators should be evaluated.  In Figure~\ref{fig:setminus}, the red curve blocks a sub tree $T'$ (\circled{1} in  Figure~\ref{fig:setminus}) whose root is set minus operator $L-R$. For the sub trees in the left and right branch of operator $L-R$, i.e., $T_1$ and $T_2$ in Figure~\ref{fig:setminus}, we apply QUIP normally on them to execute the query processing and missing value imputation. 
In addition, in operator $L-R$, all missing values in $L$ and $R$ will be imputed to make sure that the tuples in set minus operator $L-R$ can be evaluated immediately. \footnote{If there are projection operators in the top of $T_1$ and $T_2$, then all missing values in $L$ and $R$ will normally be imputed.} Except tree $T'$, QUIP will normally execute its functionality in other parts of the query tree, as shown in \circled{2} in Figure~\ref{fig:setminus}. 
The reasons to eagerly evaluate $L-R$ instead of \textit{delaying missing values in set minus operator} are two-folds. First, if there are missing values in $R$ and we delay imputations, this will cause the tuples in $L$ to be delayed for evaluation because of the uncertainty in $R$. That is, we will not know if a tuple $t\in L$ is \textit{valid} or not.\footnote{A tuple is valid if it satisfies all the predicates that are applicable in it.} If such tuples in $L$ are pushed above for later query processing, more tuples joined with $L$ will also not be valid, leading to \textit{cascade evaluations}. Thus we choose to impute all missing values in $R$ before evaluation. Similarly, if there are no missing values in $R$ but $L$ has missing values, and delay imputations in for missing values in $L$, then tuples in $L$ will also not be valid, leading to cascade evaluations, which could potentially downgrade the query performance significantly.    

\begin{figure}
\raggedright
\texttt{
\hspace{-0.5em}\textcolor{blue}{SELECT} U.name, T.time, T. room$\_$location \\ \textcolor{blue}{FROM} Trajectories as T, User as U \\
\textcolor{blue}{WHERE}
U.mac$\_$address = T.mac$\_$address
\textcolor{blue}{AND} \\
T.Room$\_$location is in \\
\{\textcolor{blue}{SELECT} S.room from Space as S\\
\textcolor{blue}{WHERE} S.Building = 'DBH'\}
}
    \caption{Nested Query}
    \label{fig:nestedquery}
\end{figure}

\begin{figure}[tb]
	\centering
	\includegraphics[width=0.5\linewidth]{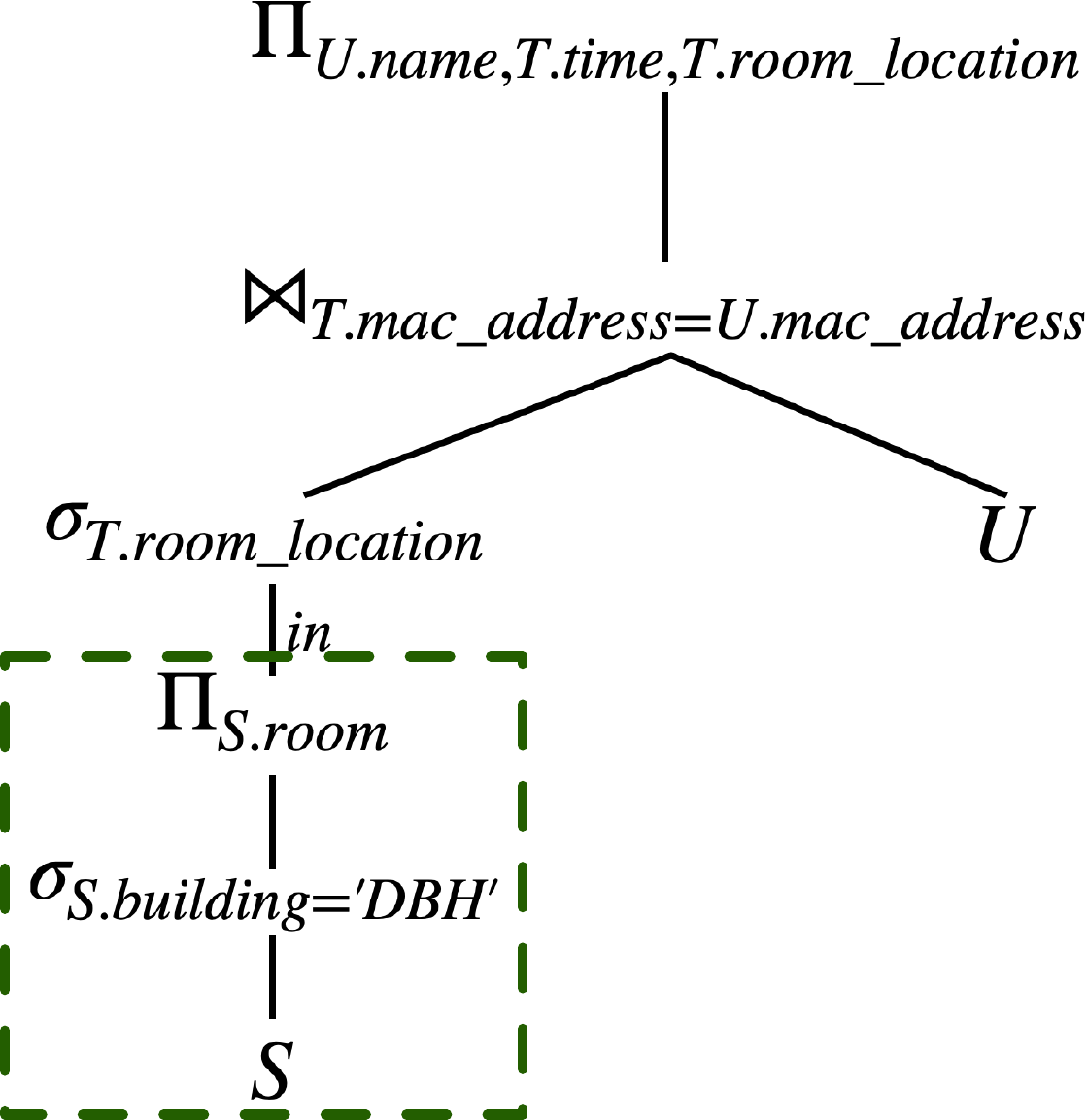}
	\caption{Nested Query Tree.}
	\label{fig:nestedquerytree}
\end{figure}

\noindent\textbf{Nested Query.} If the nested query (also known as sub query) can be rewritten as flat query, then QUIP will be normally applied in new query tree without nested query. Otherwise, we apply QUIP in the sub tree corresponding to the nested query such that the output of nested query does not contain missing values in the attributes that the outside operator directly operates on. For example, consider the nested query in Figure~\ref{fig:nestedquery}. 
Its query tree is shown in  Figure~\ref{fig:nestedquerytree}. For the sub tree $T'$ that corresponds to the green box in Figure~\ref{fig:nestedquerytree}, we call QUIP to  handle the query processing and missing value imputations there such that there are no missing values in the query answer $S.room$ returned by this sub query. Note that in this query, the sub tree $T'$ is blocking since we need to wait all satisfied $S.room$ are returned and then values in $T.Room$\_$location$ can be evaluated. For the other part of query tree, QUIP will be executed as normal.

\subsection{Evaluation}
\label{subsec:evaluation}

\subsubsection{Data Sets}
\label{subsubsec:dataset}
We report the metadata of two real data sets, \textsf{CDC} and \textsf{UCI-WiFi}, in Table~\ref{tab:cdc} and Table~\ref{tab:uci-wifi}, respectively. In particular, we show the schema and size (cardinality) of all relations in these data sets,  as well as the percentage of missing values for each attribute.

\subsubsection{Optimizations for Min/MAX Queries.}
\label{subsubsec:minmax}

In Table~\ref{tab:maxmin} we report the effect of MAX/MIN optimizations in Section~\ref{subsec:extensions} in real data sets using KNN imputations, and we denote QUIP and QUIP- as the QUIP with and without MAX/MIN optimization, and $|RT|$ as the number of tuples removed by the extra selection operators introduced by this optimization technique. In most cases, the MAX/MIN optimization helps eliminate tuples aggressively and reduce the number of imputations by around 50\% to 90\%.  It thus also speed ups  the overall query running time  considerably by around 2x to 4x in different queries. This optimization does not work well for  CDC-Q6 because we observe that the introduced selection operator is in the rightmost leaf node in the query tree, which is a left deep tree. In this case, this extra selection operator is too close to the root and thus takes effect in the very end of pipeline processing, which will not improve the performance much. 

\subsubsection{Comparison with QuERy.}
\begin{table}[]
\small
\centering
\begin{tabular}{ |c|c|c|c|c|c| } 
\hline
 & Approach & CDC & UCI-WiFi & SM-low & SM-high \\
\hline
\multirow{3}{2em}{$T$(ms)} 
& QUIP & 201.6 & 997.7 & 800.2 & 9568.3 \\ 
& QuERy-Adaptive & 441.4 & 8056.6 & 8713.5 & 17221.4 \\ 
& QuERy-Lazy & 1 min & 5 min & 43 min & > 5h\\ 
\hline
\multirow{3}{2.5em}{$T_Q$(ms)} 
& QUIP & 123.1 & 748.6 & 1280.7 & 3247.3 \\ 
& QuERy-Adaptive & 131.2 & 1149.6 & 1673.5 & 10418.7 \\ 
& QuERy-Lazy & 1 min & 5 min & 43 min & > 5h\\ 
\hline
\multirow{3}{2em}{\#Imp} 
& QUIP & 771 & 3559 & 773 & 1.81 $\times10^5$ \\ 
& QuERy-Adaptive & 4432 & 98672 & 1.72 $\times10^5$ & 1.83 $\times10^5$ \\ 
& QuERy-Lazy & 735 & 3219 & 768 & 1.79 $\times10^5$\\ 
\hline
\end{tabular}
\caption{Comparing with QuERy.}
\vspace{-1em}
    \label{tab:Cquery}
\end{table}

We compare QUIP with QuERy~\cite{altwaijry2015query} in Table~\ref{tab:Cquery}. We first denote by $T$ the total query execution time, which contains imputation costs and query processing time $T_Q$, and we denote by $\#Imp$ the number of imputations performed.  
QuERy, that is designed for ER problem,  will execute cartesian product when one of the join inputs contains dirty data. (blocks in ER problem or missing values in imputation problem) We consider two versions of QuERy and convert them to apply on the imputation problem. One is an adaptive solution, denoted by \textit{QuERy-Adaptive} which adaptively cleans the dirty data based on q cost-based solution. We modify it to apply to imputation problem by first converting the join operation in QUIP from outer join to cartesian product~\footnote{Cartesian product is executed when missing values are seen under join attributes, otherwise, inner join will be applied.} and then replacing implementation of decision function with sampling techniques in QuERy. The other is the lazy approach, denoted by \textit{QuERy-Lazy}, which always delays cleaning to the very end of query processing. This can be achieved by disabling decision function and making it always return false to implement the lazy imputation strategy. We observe that in the CDC data set, where there are no missing values under join attributes, QuERy-Adaptive has similar query processing time $T_Q$ to QUIP. However, more imputations in sampling phase by QuERy introduce extra overall overhead. In other data sets where missing values exist in join attributes, QUIP very significantly outperforms QuERy-Adaptive because the cost of join implementation by QuERy will push decision function to impute more missing values as early as possible, and thus the imputation costs increases. As for QuERy-Lazy approach, it achieves similar imputations as QUIP, but the costly join approach adds overheads significantly. 

\balance
\bibliographystyle{ACM-Reference-Format}
\bibliography{refs}


\begin{thebibliography}{44}


\ifx \showCODEN    \undefined \def \showCODEN     #1{\unskip}     \fi
\ifx \showDOI      \undefined \def \showDOI       #1{#1}\fi
\ifx \showISBNx    \undefined \def \showISBNx     #1{\unskip}     \fi
\ifx \showISBNxiii \undefined \def \showISBNxiii  #1{\unskip}     \fi
\ifx \showISSN     \undefined \def \showISSN      #1{\unskip}     \fi
\ifx \showLCCN     \undefined \def \showLCCN      #1{\unskip}     \fi
\ifx \shownote     \undefined \def \shownote      #1{#1}          \fi
\ifx \showarticletitle \undefined \def \showarticletitle #1{#1}   \fi
\ifx \showURL      \undefined \def \showURL       {\relax}        \fi
\providecommand\bibfield[2]{#2}
\providecommand\bibinfo[2]{#2}
\providecommand\natexlab[1]{#1}
\providecommand\showeprint[2][]{arXiv:#2}

\bibitem[\protect\citeauthoryear{??}{cdc}{2014}]%
        {cdcdata}
 \bibinfo{year}{2013-2014}\natexlab{}.
\newblock \bibinfo{booktitle}{\emph{Center for Disease Control. National Health
  and Nutrition Examination Survey. {\url{https://wwwn.cdc. gov / nchs / nhanes
  / ContinuousNhanes / Default . aspx ? BeginYear=2013.}}}}
\newblock


\bibitem[\protect\citeauthoryear{??}{Imp}{2017}]%
        {ImputeDB}
 \bibinfo{year}{2017}\natexlab{}.
\newblock \bibinfo{booktitle}{\emph{Github Codebase of ImputeDB.
  {\url{https://github.com/mitdbg/imputedb.git}}}}.
\newblock


\bibitem[\protect\citeauthoryear{??}{sim}{2019}]%
        {simpledb}
 \bibinfo{year}{2019}\natexlab{}.
\newblock \bibinfo{booktitle}{\emph{6.830 Lab 1: SimpleDB.
  {\url{http://db.csail.mit.edu/6. 830/assignments/lab1.html}}}}.
\newblock


\bibitem[\protect\citeauthoryear{??}{QUI}{2021}]%
        {QUIP}
 \bibinfo{year}{2021}\natexlab{}.
\newblock \bibinfo{booktitle}{\emph{Github Codebase of Quip.
  {\url{https://github.com/yiminl18/QDMIDB.git}}}}.
\newblock


\bibitem[\protect\citeauthoryear{??}{his}{2021}]%
        {histogramboost}
 \bibinfo{year}{2021}\natexlab{}.
\newblock \bibinfo{booktitle}{\emph{Histogram-Based Gradient Boosting
  Ensembles.
  {\url{https://machinelearningmastery.com/histogram-based-gradient-boosting-ensembles/}}}}.
\newblock


\bibitem[\protect\citeauthoryear{??}{blo}{2021}]%
        {bloomfilter}
 \bibinfo{year}{2021}\natexlab{}.
\newblock
  \bibinfo{booktitle}{\emph{{\url{https://en.wikipedia.org/wiki/Bloom_filter}}}}.
\newblock


\bibitem[\protect\citeauthoryear{??}{knn}{2022}]%
        {knn}
 \bibinfo{year}{2022}\natexlab{}.
\newblock
  \bibinfo{booktitle}{\emph{{\url{https://scikit-learn.org/stable/modules/generated/sklearn.impute.KNNImputer.html}}}}.
\newblock


\bibitem[\protect\citeauthoryear{??}{pos}{2022}]%
        {postgresql}
 \bibinfo{year}{2022}\natexlab{}.
\newblock \bibinfo{booktitle}{\emph{PostgreSQL Index and Order.
  {\url{https://www.postgresql.org/docs/9.1/indexes-ordering.html}}}}.
\newblock


\bibitem[\protect\citeauthoryear{??}{uci}{2022}]%
        {ucioccupancy}
 \bibinfo{year}{2022}\natexlab{}.
\newblock \bibinfo{booktitle}{\emph{UCI Real-time Occupancy Tracing
  Application.
  {\url{https://tippersweb.ics.uci.edu/covid19/d/UUeKIMPMz/engineering-occupancy-counts?orgId=1&from=now-24h&to=now}}}}.
\newblock


\bibitem[\protect\citeauthoryear{Altwaijry, Kalashnikov, and
  Mehrotra}{Altwaijry et~al\mbox{.}}{2013}]%
        {altwaijry2013query}
\bibfield{author}{\bibinfo{person}{Hotham Altwaijry}, \bibinfo{person}{Dmitri~V
  Kalashnikov}, {and} \bibinfo{person}{Sharad Mehrotra}.}
  \bibinfo{year}{2013}\natexlab{}.
\newblock \showarticletitle{Query-driven approach to entity resolution}.
\newblock \bibinfo{journal}{\emph{Proceedings of the VLDB Endowment}}
  \bibinfo{volume}{6}, \bibinfo{number}{14} (\bibinfo{year}{2013}),
  \bibinfo{pages}{1846--1857}.
\newblock


\bibitem[\protect\citeauthoryear{Altwaijry, Mehrotra, and
  Kalashnikov}{Altwaijry et~al\mbox{.}}{2015}]%
        {altwaijry2015query}
\bibfield{author}{\bibinfo{person}{Hotham Altwaijry}, \bibinfo{person}{Sharad
  Mehrotra}, {and} \bibinfo{person}{Dmitri~V Kalashnikov}.}
  \bibinfo{year}{2015}\natexlab{}.
\newblock \showarticletitle{Query: A framework for integrating entity
  resolution with query processing}.
\newblock \bibinfo{journal}{\emph{Proceedings of the VLDB Endowment}}
  \bibinfo{volume}{9}, \bibinfo{number}{3} (\bibinfo{year}{2015}),
  \bibinfo{pages}{120--131}.
\newblock


\bibitem[\protect\citeauthoryear{Arjovsky, Chintala, and Bottou}{Arjovsky
  et~al\mbox{.}}{2017}]%
        {arjovsky2017wasserstein}
\bibfield{author}{\bibinfo{person}{Martin Arjovsky}, \bibinfo{person}{Soumith
  Chintala}, {and} \bibinfo{person}{L{\'e}on Bottou}.}
  \bibinfo{year}{2017}\natexlab{}.
\newblock \showarticletitle{Wasserstein generative adversarial networks}. In
  \bibinfo{booktitle}{\emph{International conference on machine learning}}.
  PMLR, \bibinfo{pages}{214--223}.
\newblock


\bibitem[\protect\citeauthoryear{Athanassoulis, B{\o}gh, and
  Idreos}{Athanassoulis et~al\mbox{.}}{2019}]%
        {athanassoulis2019optimal}
\bibfield{author}{\bibinfo{person}{Manos Athanassoulis},
  \bibinfo{person}{Kenneth~S B{\o}gh}, {and} \bibinfo{person}{Stratos Idreos}.}
  \bibinfo{year}{2019}\natexlab{}.
\newblock \showarticletitle{Optimal column layout for hybrid workloads}.
\newblock \bibinfo{journal}{\emph{Proceedings of the VLDB Endowment}}
  \bibinfo{volume}{12}, \bibinfo{number}{13} (\bibinfo{year}{2019}),
  \bibinfo{pages}{2393--2407}.
\newblock


\bibitem[\protect\citeauthoryear{Bansal, Deshpande, and Sarawagi}{Bansal
  et~al\mbox{.}}{2021}]%
        {bansal2021missing}
\bibfield{author}{\bibinfo{person}{Parikshit Bansal},
  \bibinfo{person}{Prathamesh Deshpande}, {and} \bibinfo{person}{Sunita
  Sarawagi}.} \bibinfo{year}{2021}\natexlab{}.
\newblock \showarticletitle{Missing value imputation on multidimensional time
  series}.
\newblock \bibinfo{journal}{\emph{arXiv preprint arXiv:2103.01600}}
  (\bibinfo{year}{2021}).
\newblock


\bibitem[\protect\citeauthoryear{Bilenko, Basil, and Sahami}{Bilenko
  et~al\mbox{.}}{2005}]%
        {bilenko2005adaptive}
\bibfield{author}{\bibinfo{person}{Mikhail Bilenko}, \bibinfo{person}{S Basil},
  {and} \bibinfo{person}{Mehran Sahami}.} \bibinfo{year}{2005}\natexlab{}.
\newblock \showarticletitle{Adaptive product normalization: Using online
  learning for record linkage in comparison shopping}. In
  \bibinfo{booktitle}{\emph{Fifth IEEE International Conference on Data Mining
  (ICDM'05)}}. IEEE, \bibinfo{pages}{8--pp}.
\newblock


\bibitem[\protect\citeauthoryear{Burgette and Reiter}{Burgette and
  Reiter}{2010}]%
        {burgette2010multiple}
\bibfield{author}{\bibinfo{person}{Lane~F Burgette} {and}
  \bibinfo{person}{Jerome~P Reiter}.} \bibinfo{year}{2010}\natexlab{}.
\newblock \showarticletitle{Multiple imputation for missing data via sequential
  regression trees}.
\newblock \bibinfo{journal}{\emph{American journal of epidemiology}}
  \bibinfo{volume}{172}, \bibinfo{number}{9} (\bibinfo{year}{2010}),
  \bibinfo{pages}{1070--1076}.
\newblock


\bibitem[\protect\citeauthoryear{Cambronero, Feser, Smith, and
  Madden}{Cambronero et~al\mbox{.}}{2017}]%
        {cambronero2017query}
\bibfield{author}{\bibinfo{person}{Jos{\'e} Cambronero},
  \bibinfo{person}{John~K Feser}, \bibinfo{person}{Micah~J Smith}, {and}
  \bibinfo{person}{Samuel Madden}.} \bibinfo{year}{2017}\natexlab{}.
\newblock \showarticletitle{Query optimization for dynamic imputation}.
\newblock \bibinfo{journal}{\emph{Proceedings of the VLDB Endowment}}
  \bibinfo{volume}{10}, \bibinfo{number}{11} (\bibinfo{year}{2017}),
  \bibinfo{pages}{1310--1321}.
\newblock


\bibitem[\protect\citeauthoryear{Chen and Guestrin}{Chen and Guestrin}{2016}]%
        {chen2016xgboost}
\bibfield{author}{\bibinfo{person}{Tianqi Chen} {and} \bibinfo{person}{Carlos
  Guestrin}.} \bibinfo{year}{2016}\natexlab{}.
\newblock \showarticletitle{Xgboost: A scalable tree boosting system}. In
  \bibinfo{booktitle}{\emph{Proceedings of the 22nd acm sigkdd international
  conference on knowledge discovery and data mining}}.
  \bibinfo{pages}{785--794}.
\newblock


\bibitem[\protect\citeauthoryear{Chu, Morcos, Ilyas, Ouzzani, Papotti, Tang,
  and Ye}{Chu et~al\mbox{.}}{2015}]%
        {chu2015katara}
\bibfield{author}{\bibinfo{person}{Xu Chu}, \bibinfo{person}{John Morcos},
  \bibinfo{person}{Ihab~F Ilyas}, \bibinfo{person}{Mourad Ouzzani},
  \bibinfo{person}{Paolo Papotti}, \bibinfo{person}{Nan Tang}, {and}
  \bibinfo{person}{Yin Ye}.} \bibinfo{year}{2015}\natexlab{}.
\newblock \showarticletitle{Katara: A data cleaning system powered by knowledge
  bases and crowdsourcing}. In \bibinfo{booktitle}{\emph{Proceedings of the
  2015 ACM SIGMOD international conference on management of data}}.
  \bibinfo{pages}{1247--1261}.
\newblock


\bibitem[\protect\citeauthoryear{Fan, Li, Ma, Tang, and Yu}{Fan
  et~al\mbox{.}}{2012}]%
        {fan2012towards}
\bibfield{author}{\bibinfo{person}{Wenfei Fan}, \bibinfo{person}{Jianzhong Li},
  \bibinfo{person}{Shuai Ma}, \bibinfo{person}{Nan Tang}, {and}
  \bibinfo{person}{Wenyuan Yu}.} \bibinfo{year}{2012}\natexlab{}.
\newblock \showarticletitle{Towards certain fixes with editing rules and master
  data}.
\newblock \bibinfo{journal}{\emph{The VLDB journal}} \bibinfo{volume}{21},
  \bibinfo{number}{2} (\bibinfo{year}{2012}), \bibinfo{pages}{213--238}.
\newblock


\bibitem[\protect\citeauthoryear{Giannakopoulou, Karpathiotakis, and
  Ailamaki}{Giannakopoulou et~al\mbox{.}}{2020}]%
        {giannakopoulou2020cleaning}
\bibfield{author}{\bibinfo{person}{Stella Giannakopoulou},
  \bibinfo{person}{Manos Karpathiotakis}, {and} \bibinfo{person}{Anastasia
  Ailamaki}.} \bibinfo{year}{2020}\natexlab{}.
\newblock \showarticletitle{Cleaning Denial Constraint Violations through
  Relaxation}. In \bibinfo{booktitle}{\emph{Proceedings of the 2020 ACM SIGMOD
  International Conference on Management of Data}}. \bibinfo{pages}{805--815}.
\newblock


\bibitem[\protect\citeauthoryear{Graham}{Graham}{2009}]%
        {graham2009missing}
\bibfield{author}{\bibinfo{person}{John~W Graham}.}
  \bibinfo{year}{2009}\natexlab{}.
\newblock \showarticletitle{Missing data analysis: Making it work in the real
  world}.
\newblock \bibinfo{journal}{\emph{Annual review of psychology}}
  \bibinfo{volume}{60} (\bibinfo{year}{2009}), \bibinfo{pages}{549--576}.
\newblock


\bibitem[\protect\citeauthoryear{Gupta, Carey, Mehrotra, and Yus}{Gupta
  et~al\mbox{.}}{2020}]%
        {gupta2020smartbench}
\bibfield{author}{\bibinfo{person}{Peeyush Gupta}, \bibinfo{person}{Michael~J
  Carey}, \bibinfo{person}{Sharad Mehrotra}, {and} \bibinfo{person}{oberto
  Yus}.} \bibinfo{year}{2020}\natexlab{}.
\newblock \showarticletitle{SmartBench: a benchmark for data management in
  smart spaces}.
\newblock \bibinfo{journal}{\emph{Proceedings of the VLDB Endowment}}
  \bibinfo{volume}{13}, \bibinfo{number}{12} (\bibinfo{year}{2020}),
  \bibinfo{pages}{1807--1820}.
\newblock


\bibitem[\protect\citeauthoryear{Herschel, Hern{\'a}ndez, and Tan}{Herschel
  et~al\mbox{.}}{2009}]%
        {herschel2009artemis}
\bibfield{author}{\bibinfo{person}{Melanie Herschel},
  \bibinfo{person}{Mauricio~A Hern{\'a}ndez}, {and} \bibinfo{person}{Wang-Chiew
  Tan}.} \bibinfo{year}{2009}\natexlab{}.
\newblock \showarticletitle{Artemis: A system for analyzing missing answers}.
\newblock \bibinfo{journal}{\emph{Proceedings of the VLDB Endowment}}
  \bibinfo{volume}{2}, \bibinfo{number}{2} (\bibinfo{year}{2009}),
  \bibinfo{pages}{1550--1553}.
\newblock


\bibitem[\protect\citeauthoryear{Ioannou, Nejdl, Nieder{\'e}e, and
  Velegrakis}{Ioannou et~al\mbox{.}}{2010}]%
        {ioannou2010fly}
\bibfield{author}{\bibinfo{person}{Ekaterini Ioannou},
  \bibinfo{person}{Wolfgang Nejdl}, \bibinfo{person}{Claudia Nieder{\'e}e},
  {and} \bibinfo{person}{Yannis Velegrakis}.} \bibinfo{year}{2010}\natexlab{}.
\newblock \showarticletitle{On-the-fly entity-aware query processing in the
  presence of linkage}.
\newblock \bibinfo{journal}{\emph{Proceedings of the VLDB Endowment}}
  \bibinfo{volume}{3}, \bibinfo{number}{1-2} (\bibinfo{year}{2010}),
  \bibinfo{pages}{429--438}.
\newblock


\bibitem[\protect\citeauthoryear{Jagadish, Koudas, Muthukrishnan, Poosala,
  Sevcik, and Suel}{Jagadish et~al\mbox{.}}{1998}]%
        {jagadish1998optimal}
\bibfield{author}{\bibinfo{person}{Hosagrahar~Visvesvaraya Jagadish},
  \bibinfo{person}{Nick Koudas}, \bibinfo{person}{S Muthukrishnan},
  \bibinfo{person}{Viswanath Poosala}, \bibinfo{person}{Kenneth~C Sevcik},
  {and} \bibinfo{person}{Torsten Suel}.} \bibinfo{year}{1998}\natexlab{}.
\newblock \showarticletitle{Optimal histograms with quality guarantees}. In
  \bibinfo{booktitle}{\emph{VLDB}}, Vol.~\bibinfo{volume}{98}.
  \bibinfo{pages}{24--27}.
\newblock


\bibitem[\protect\citeauthoryear{Ke, Meng, Finley, Wang, Chen, Ma, Ye, and
  Liu}{Ke et~al\mbox{.}}{2017}]%
        {ke2017lightgbm}
\bibfield{author}{\bibinfo{person}{Guolin Ke}, \bibinfo{person}{Qi Meng},
  \bibinfo{person}{Thomas Finley}, \bibinfo{person}{Taifeng Wang},
  \bibinfo{person}{Wei Chen}, \bibinfo{person}{Weidong Ma},
  \bibinfo{person}{Qiwei Ye}, {and} \bibinfo{person}{Tie-Yan Liu}.}
  \bibinfo{year}{2017}\natexlab{}.
\newblock \showarticletitle{Lightgbm: A highly efficient gradient boosting
  decision tree}.
\newblock \bibinfo{journal}{\emph{Advances in neural information processing
  systems}}  \bibinfo{volume}{30} (\bibinfo{year}{2017}).
\newblock


\bibitem[\protect\citeauthoryear{Khayati, Lerner, Tymchenko, and
  Cudr{\'e}-Mauroux}{Khayati et~al\mbox{.}}{2020}]%
        {khayati2020mind}
\bibfield{author}{\bibinfo{person}{Mourad Khayati}, \bibinfo{person}{Alberto
  Lerner}, \bibinfo{person}{Zakhar Tymchenko}, {and} \bibinfo{person}{Philippe
  Cudr{\'e}-Mauroux}.} \bibinfo{year}{2020}\natexlab{}.
\newblock \showarticletitle{Mind the gap: an experimental evaluation of
  imputation of missing values techniques in time series}. In
  \bibinfo{booktitle}{\emph{Proceedings of the VLDB Endowment}},
  Vol.~\bibinfo{volume}{13}. \bibinfo{pages}{768--782}.
\newblock


\bibitem[\protect\citeauthoryear{Kim, Choi, Hong, Kim, and Lee}{Kim
  et~al\mbox{.}}{2003}]%
        {kim2003taxonomy}
\bibfield{author}{\bibinfo{person}{Won Kim}, \bibinfo{person}{Byoung-Ju Choi},
  \bibinfo{person}{Eui-Kyeong Hong}, \bibinfo{person}{Soo-Kyung Kim}, {and}
  \bibinfo{person}{Doheon Lee}.} \bibinfo{year}{2003}\natexlab{}.
\newblock \showarticletitle{A taxonomy of dirty data}.
\newblock \bibinfo{journal}{\emph{Data mining and knowledge discovery}}
  \bibinfo{volume}{7}, \bibinfo{number}{1} (\bibinfo{year}{2003}),
  \bibinfo{pages}{81--99}.
\newblock


\bibitem[\protect\citeauthoryear{Li, Rao, Blase, Zhang, Chu, and Zhang}{Li
  et~al\mbox{.}}{2021}]%
        {li2021cleanml}
\bibfield{author}{\bibinfo{person}{Peng Li}, \bibinfo{person}{Xi Rao},
  \bibinfo{person}{Jennifer Blase}, \bibinfo{person}{Yue Zhang},
  \bibinfo{person}{Xu Chu}, {and} \bibinfo{person}{Ce Zhang}.}
  \bibinfo{year}{2021}\natexlab{}.
\newblock \showarticletitle{CleanML: a study for evaluating the impact of data
  cleaning on ml classification tasks}. In \bibinfo{booktitle}{\emph{36th IEEE
  International Conference on Data Engineering (ICDE 2020)(virtual)}}. ETH
  Zurich, Institute for Computing Platforms.
\newblock


\bibitem[\protect\citeauthoryear{Lin and Tsai}{Lin and Tsai}{2020}]%
        {lin2020missing}
\bibfield{author}{\bibinfo{person}{Wei-Chao Lin} {and}
  \bibinfo{person}{Chih-Fong Tsai}.} \bibinfo{year}{2020}\natexlab{}.
\newblock \showarticletitle{Missing value imputation: a review and analysis of
  the literature (2006--2017)}.
\newblock \bibinfo{journal}{\emph{Artificial Intelligence Review}}
  \bibinfo{volume}{53}, \bibinfo{number}{2} (\bibinfo{year}{2020}),
  \bibinfo{pages}{1487--1509}.
\newblock


\bibitem[\protect\citeauthoryear{Lin et~al\mbox{.}}{Lin et~al\mbox{.}}{2021a}]%
        {lin2020locater}
\bibfield{author}{\bibinfo{person}{Yiming Lin} {et~al\mbox{.}}}
  \bibinfo{year}{2021}\natexlab{a}.
\newblock \showarticletitle{Locater: cleaning wifi connectivity datasets for
  semantic localization}.
\newblock \bibinfo{journal}{\emph{Proceedings of the VLDB Endowment}}
  \bibinfo{number}{3} (\bibinfo{year}{2021}), \bibinfo{pages}{329 -- 341}.
\newblock


\bibitem[\protect\citeauthoryear{Lin, Khargonekar, Mehrotra, and
  Venkatasubramanian}{Lin et~al\mbox{.}}{2021b}]%
        {lin2021t}
\bibfield{author}{\bibinfo{person}{Yiming Lin}, \bibinfo{person}{Pramod
  Khargonekar}, \bibinfo{person}{Sharad Mehrotra}, {and}
  \bibinfo{person}{Nalini Venkatasubramanian}.}
  \bibinfo{year}{2021}\natexlab{b}.
\newblock \showarticletitle{T-cove: an exposure tracing system based on
  cleaning wi-fi events on organizational premises}.
\newblock \bibinfo{journal}{\emph{Proceedings of the VLDB Endowment}}
  \bibinfo{volume}{14}, \bibinfo{number}{12} (\bibinfo{year}{2021}),
  \bibinfo{pages}{2783--2786}.
\newblock


\bibitem[\protect\citeauthoryear{Makridakis and Hibon}{Makridakis and
  Hibon}{2000}]%
        {makridakis2000m3}
\bibfield{author}{\bibinfo{person}{Spyros Makridakis} {and}
  \bibinfo{person}{Michele Hibon}.} \bibinfo{year}{2000}\natexlab{}.
\newblock \showarticletitle{The M3-Competition: results, conclusions and
  implications}.
\newblock \bibinfo{journal}{\emph{International journal of forecasting}}
  \bibinfo{volume}{16}, \bibinfo{number}{4} (\bibinfo{year}{2000}),
  \bibinfo{pages}{451--476}.
\newblock


\bibitem[\protect\citeauthoryear{McCarthy and Dayal}{McCarthy and
  Dayal}{1989}]%
        {mccarthy1989architecture}
\bibfield{author}{\bibinfo{person}{Dennis McCarthy} {and}
  \bibinfo{person}{Umeshwar Dayal}.} \bibinfo{year}{1989}\natexlab{}.
\newblock \showarticletitle{The architecture of an active database management
  system}.
\newblock \bibinfo{journal}{\emph{ACM Sigmod Record}} \bibinfo{volume}{18},
  \bibinfo{number}{2} (\bibinfo{year}{1989}), \bibinfo{pages}{215--224}.
\newblock


\bibitem[\protect\citeauthoryear{Mehrotra et~al\mbox{.}}{Mehrotra
  et~al\mbox{.}}{2016}]%
        {mehrotra2016tippers}
\bibfield{author}{\bibinfo{person}{Sharad Mehrotra} {et~al\mbox{.}}}
  \bibinfo{year}{2016}\natexlab{}.
\newblock \showarticletitle{TIPPERS: A privacy cognizant IoT environment}. In
  \bibinfo{booktitle}{\emph{PerCom Workshops}}. \bibinfo{pages}{1--6}.
\newblock


\bibitem[\protect\citeauthoryear{Miao, Wu, Chen, Gao, Wang, and Yin}{Miao
  et~al\mbox{.}}{2021}]%
        {miao2021efficient}
\bibfield{author}{\bibinfo{person}{Xiaoye Miao}, \bibinfo{person}{Yangyang Wu},
  \bibinfo{person}{Lu Chen}, \bibinfo{person}{Yunjun Gao}, \bibinfo{person}{Jun
  Wang}, {and} \bibinfo{person}{Jianwei Yin}.} \bibinfo{year}{2021}\natexlab{}.
\newblock \showarticletitle{Efficient and effective data imputation with
  influence functions}.
\newblock \bibinfo{journal}{\emph{Proceedings of the VLDB Endowment}}
  \bibinfo{volume}{15}, \bibinfo{number}{3} (\bibinfo{year}{2021}),
  \bibinfo{pages}{624--632}.
\newblock


\bibitem[\protect\citeauthoryear{Qi, Wang, Li, and Gao}{Qi
  et~al\mbox{.}}{2018}]%
        {qi2018frog}
\bibfield{author}{\bibinfo{person}{Zhixin Qi}, \bibinfo{person}{Hongzhi Wang},
  \bibinfo{person}{Jianzhong Li}, {and} \bibinfo{person}{Hong Gao}.}
  \bibinfo{year}{2018}\natexlab{}.
\newblock \showarticletitle{FROG: Inference from knowledge base for missing
  value imputation}.
\newblock \bibinfo{journal}{\emph{Knowledge-Based Systems}}
  \bibinfo{volume}{145} (\bibinfo{year}{2018}), \bibinfo{pages}{77--90}.
\newblock


\bibitem[\protect\citeauthoryear{Song, Zhang, Chen, and Wang}{Song
  et~al\mbox{.}}{2015}]%
        {song2015enriching}
\bibfield{author}{\bibinfo{person}{Shaoxu Song}, \bibinfo{person}{Aoqian
  Zhang}, \bibinfo{person}{Lei Chen}, {and} \bibinfo{person}{Jianmin Wang}.}
  \bibinfo{year}{2015}\natexlab{}.
\newblock \showarticletitle{Enriching data imputation with extensive similarity
  neighbors}.
\newblock \bibinfo{journal}{\emph{Proceedings of the VLDB Endowment}}
  \bibinfo{volume}{8}, \bibinfo{number}{11} (\bibinfo{year}{2015}),
  \bibinfo{pages}{1286--1297}.
\newblock


\bibitem[\protect\citeauthoryear{Tan and Goh}{Tan and Goh}{1999}]%
        {tan1999implementing}
\bibfield{author}{\bibinfo{person}{CW Tan} {and} \bibinfo{person}{Angela Goh}.}
  \bibinfo{year}{1999}\natexlab{}.
\newblock \showarticletitle{Implementing ECA rules in an active database}.
\newblock \bibinfo{journal}{\emph{Knowledge-Based Systems}}
  \bibinfo{volume}{12}, \bibinfo{number}{4} (\bibinfo{year}{1999}),
  \bibinfo{pages}{137--144}.
\newblock


\bibitem[\protect\citeauthoryear{Thaper, Guha, Indyk, and Koudas}{Thaper
  et~al\mbox{.}}{2002}]%
        {thaper2002dynamic}
\bibfield{author}{\bibinfo{person}{Nitin Thaper}, \bibinfo{person}{Sudipto
  Guha}, \bibinfo{person}{Piotr Indyk}, {and} \bibinfo{person}{Nick Koudas}.}
  \bibinfo{year}{2002}\natexlab{}.
\newblock \showarticletitle{Dynamic multidimensional histograms}. In
  \bibinfo{booktitle}{\emph{Proceedings of the 2002 ACM SIGMOD international
  conference on Management of data}}. \bibinfo{pages}{428--439}.
\newblock


\bibitem[\protect\citeauthoryear{Ye, Wang, Li, Gao, and Cheng}{Ye
  et~al\mbox{.}}{2016}]%
        {ye2016crowdsourcing}
\bibfield{author}{\bibinfo{person}{Chen Ye}, \bibinfo{person}{Hongzhi Wang},
  \bibinfo{person}{Jianzhong Li}, \bibinfo{person}{Hong Gao}, {and}
  \bibinfo{person}{Siyao Cheng}.} \bibinfo{year}{2016}\natexlab{}.
\newblock \showarticletitle{Crowdsourcing-enhanced missing values imputation
  based on Bayesian network}. In \bibinfo{booktitle}{\emph{International
  Conference on Database Systems for Advanced Applications}}. Springer,
  \bibinfo{pages}{67--81}.
\newblock


\bibitem[\protect\citeauthoryear{Ye, Wang, Lu, and Li}{Ye
  et~al\mbox{.}}{2020}]%
        {ye2020effective}
\bibfield{author}{\bibinfo{person}{Chen Ye}, \bibinfo{person}{Hongzhi Wang},
  \bibinfo{person}{Wenbo Lu}, {and} \bibinfo{person}{Jianzhong Li}.}
  \bibinfo{year}{2020}\natexlab{}.
\newblock \showarticletitle{Effective Bayesian-network-based missing value
  imputation enhanced by crowdsourcing}.
\newblock \bibinfo{journal}{\emph{Knowledge-Based Systems}}
  \bibinfo{volume}{190} (\bibinfo{year}{2020}), \bibinfo{pages}{105199}.
\newblock


\bibitem[\protect\citeauthoryear{Young, Weckman, and Holland}{Young
  et~al\mbox{.}}{2011}]%
        {young2011survey}
\bibfield{author}{\bibinfo{person}{William Young}, \bibinfo{person}{Gary
  Weckman}, {and} \bibinfo{person}{W Holland}.}
  \bibinfo{year}{2011}\natexlab{}.
\newblock \showarticletitle{A survey of methodologies for the treatment of
  missing values within datasets: Limitations and benefits}.
\newblock \bibinfo{journal}{\emph{Theoretical Issues in Ergonomics Science}}
  \bibinfo{volume}{12}, \bibinfo{number}{1} (\bibinfo{year}{2011}),
  \bibinfo{pages}{15--43}.
\newblock


\end{thebibliography}

\end{document}